\documentclass[onecolumn]{emulateapj}

\shorttitle{Polarized Emission from T Tauri Disks}
\shortauthors{Cho \& Lazarian}

\begin{document}

\title{Grain Alignment and Polarized Emission from 
Magnetized T Tauri Disks}

\author{Jungyeon Cho}
\affil{Dept. of Astronomy and Space Science, Chungnam National University, Daejeon, South Korea; jcho@cnu.ac.kr}
\author{A. Lazarian}
\affil{Astronomy Dept., Univ.~of Wisconsin, Madison,
       WI53706, USA; lazarian@astro.wisc.edu}

\begin{abstract}
The structure of magnetic fields within protostellar disks may
be studied via polarimetry provided that grains are aligned in
respect to magnetic field within the disks.
We explore alignment of dust grains by radiative torque in 
T Tauri disks and provide predictions for polarized
emission for disks viewed at different wavelengths and viewing angles. 
We show that
the alignment is especially efficient in the outer parts
of the disks.
In the presence of magnetic field, these aligned grains
produce polarized emission in infrared wavelengths.
We consider a simple model of an accretion disk and
provide predictions for polarization that are available
to the present-day instruments that do not resolve the disks and
will be available to future instruments
that will resolve the disks. As the surface magnetic field and the
bulk magnetic field play different roles for the disk dynamics,
we consider separately the contributions 
that arise from the surface areas of the disk and its interior.
We find that the polarized emission drops for wavelengths shorter
than $\sim 10 \mu m$.
Between $\sim 10 \mu m$ and $\sim 100 \mu m$, the polarized 
emission is dominated by the emission from the surface layer
of the disks and the degree of polarization can be as large as
$\sim 10\%$ for unresolved disks.
We find that the degree of polarization at these wavelengths
is very sensitive to the size distribution of
dust grains in the disk surface layer, which should allow testing
the predicted grain size distributions. 
The degree of polarization in the far-infrared/sub-millimeter
wavelengths 
is sensitive to the size distribution of
dust grains in the disk interior. 
When we take a Mathis-Rumpl-Nordsieck-type distribution 
with maximum grain size of 500-1000 $\mu m$, 
the degree of polarization is
 around 2-3 \% level at wavelengths larger than $\sim$100$\mu m$.
Our study indicates 
that multifrequency infrared polarimetric studies of protostellar disks
can provide good insights into the details of their magnetic structure.
\end{abstract}
\keywords{ accretion, accretion disks ---circumstellar matter
  --- polarization --- stars: pre-main-sequence --- dust, extinction }

\section{Introduction}
Magnetic field plays important roles in star formation, as well as
formation and evolution of protostellar disks.
Magnetic pressure can provide extra support to the disks and
magnetic field can promote removal of angular momentum from disks (see
Velikov 1959; Chandrasekhar 1961; Balbus \& Hawley 1991).
However, there are many uncertainties for the structure and effects
of the magnetic field in protostellar disks. 

Infrared (IR) polarimetry may be an important tool to
investigate magnetic field structure in protostellar disks, provided that the
grains are aligned in the disks in respect to magnetic field.
Since the grain emissivity is larger for the long axis of a grain,
emitted radiation has a polarization vector parallel to
the grain's long axis. If grains are aligned
with their long axes perpendicular to magnetic field, the
resulting grain emission has polarization directed perpendicular to the
magnetic field. Therefore, by measuring the direction of polarization of
IR emission from dust grains 
one can infer the direction of magnetic field. 
The key question thus is whether grain alignment is
efficient in protostellar disks.

The notion that the grains can be aligned in respect to magnetic field
can be traced back to the discovery of star-light polarization 
by Hall (1949) and Hiltner (1949), that arises from interstellar grains.
Historically the theory of the grain alignment was developing mostly to explain the interstellar
polarization, but grain alignment is a much wider spread phenomenon (see
Lazarian 2007 for a review). Among the alignment mechanisms the one 
related
to radiative torques (RTs) looks the most promising. We invoke it for our
calculations below.

The RTs make use of interaction of radiation with a grain 
to spin the grain up.
The RT alignment was first discussed by
Dolginov (1972) and Dolginov \& Mytrophanov (1976).
However, quantitative studies were done only in 1990's.
In their pioneering work, Draine \& Weingartner 
(1996, 1997, henceforth DW96 and DW97) demonstrated the efficiency of the
RT alignment for a few arbitrary chosen
irregular grains using numerical simulations. This work identified RTs as
potentially the major agent for interstellar grain alignment.
 Cho \& Lazarian (2005, henceforth CL05) 
demonstrated the rapid increase of radiative torque efficiency and
showed that radiative alignment can
naturally explain decrease of the degree of polarization
near the centers of pre-stellar cores. Large grains are known to be present
in protostellar disk environments
 and this makes the RT alignment promising.

The effect of RTs is two-fold. They can spin-up grains and
drive the alignment. While the details of the second process are a subject
of intensive research 
(see Weingartner \& Draine 2003; Lazarian \& Hoang 2007; Hoang \& Lazarian
2007), for our estimates we use the RTs spin-up efficiency to
evaluate the efficiency of grain alignment. As
grains of different temperatures are present in protostellar disks, 
the differential alignment of
grains at different optical depths is expected to show itself through 
variations of polarization at different wavelengths.

Protostellar disks are often detected through far-infrared excess.
Dust grains in the protostellar disks are
the main cause of the infrared excess - dust grains absorb stellar radiation
and re-emit at infrared wavelengths. 
The spectral energy distribution (SED) of the emitted light
gives valuable information about the disk structure.
Recently proposed hydrostatic, radiative equilibrium passive
disk model
(Chiang \& Goldreich 1997; Chiang et al. 2001; hereafter CG97 and C01,
 respectively)
fits observed SED from T Tauri stars very well
and seems to be one of the most promising models.
Here, passive disk means that active accretion effect, which might be 
very important in the immediate vicinity of the central star, is not 
included in the model.
In this paper we use the model in C01.

Polarization arising from disks is of great interest and importance. 
Recently Aitken et al. (2002) studied polarization that can
arise from magnetized accretion disks.
They considered a single grain component consisting of
the 0.6$\mu m$ silicate and assumed that grains were
partially aligned with $R=0.25$, 
where $R=< 3 \cos^2 \beta - 1 >/2$ is the Rayleigh reduction factor (Greenberg 1968). Here $\beta$ is the precession angle between
the grain spin axis and the magnetic field and the angle brackets
denote average (see Aitken et al. 2002 for details).
In this paper we use both a theoretically motivated model of grain alignment
and a more sophisticated model of accretion disk.

Note, that challenges of observing polarization
in mid-IR to far-IR (FIR) wavelengths have been dealt with successfully
recently. For instance, Tamura et al. (1999) first detected
polarized emission from T Tauri stars, low mass 
protostars. As technology develops,
the future of polarimetric IR and submillimeter
emission  studies looks very promising.
In this paper
we try to theoretically predict
polarized emission from T Tauri disks at different wavelengths and
for different inclinations of the disks.

We calculate grain alignment by radiative torque using 
a T Tauri disk model in C01 and we predict
polarized mid-IR/FIR/sub-millimeter emission. 
In \S2, we discuss grain alignment in T Tauri disks.
In \S3, we give theoretical estimates for degree of polarization.
In \S4, we calculate the spectral energy distribution of 
{\it maximally} polarized emission, which will be useful
only when we spatially resolve disks.
In \S5, we discuss the effect of inclination angle.
In \S6, we discuss observational implications.
We give summary in \S7.

\section{Grain alignment in protostellar disks} \label{sect:diskmodel}
\subsection{The disk model used for this study}
\begin{figure}[h!t]
\includegraphics[width=.38\textwidth]{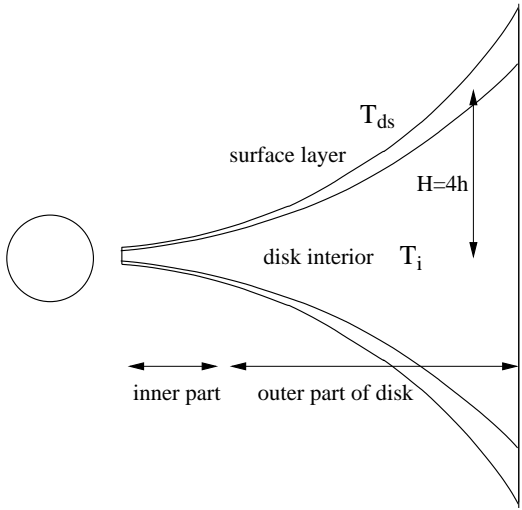}
\caption{ A schematic view of the disk model (see C01).
The surface layer is hotter and heated by the star light.
The disk interior is heated by re-processed light from the
surface layers. We assume that the disk height, $H$, is 4 time the
disk scale height, $h$.
\label{fig:model}
}
\end{figure}

 We assume that magnetic field is regular and toroidal 
(i.e. {\it azimuthal}).
We use a T Tauri disk model in C01.
Figure \ref{fig:model} schematically  shows the model.
The disk is in hydrostatic and radiative equilibrium (see also
CG97) and shows flaring.
According to CG97, flaring of disk is essential for correct
description of SED.
They considered a two-layered disk model. 
Dust grains in the surface layer
are heated directly by the radiation from the central star
and emit their heat more or less isotropically.
Half of the dust thermal emission immediately escapes and the other half
enters into disk interior and heats dusts and gas there.
Both CG97 and C01
assumed that the disk interior is isothermal.

In both CG97 and C01, the disk surface layer is
hotter than the disk interior. Thus, roughly speaking, the surface layer
dominates in mid-infrared wavelengths and disk interior
dominates in far-infrared/sub-millimeter wavelengths.
The disk surface layer is both optically and physically thin.

The major difference between CG97 and
C01 is the treatment of dust grain size
distribution.
CG97 assumed that all grains have a fixed size of
$0.1 \mu m$, while C01 assumes 
an MRN distribution (Mathis, Rumpl, \& Nordsieck 1977)
with maximum grain size of $a_{max,i} = 1000 \mu m$ in the
disk interior and $a_{max,s} = 1 \mu m$ in the disk surface layer.

In our calculations, we use a grain
model similar to that in C01.
We use an MRN-type power-law distribution of grain radii $a$ between
$a_{min}$ (=0.01 $\mu m$ for both disk interior and surface layer)
 and $a_{max}$ (=1000 $\mu m$ for disk interior and $=1\mu m$
for disk surface layer)
with a power index of -3.5: $dN \propto a^{-3.5} da$.
As in C01 we assume that grain composition varies 
with distance from the central star
in both disk interior and surface layer.
We assume that grains in the surface layer 
are made of silicate only when the distance 
$r$ is less than 6 AU, and silicate covered with water ice 
when $r>6$AU.
We do not use iron grains for the immediate vicinity of
the star.
We assume that grains in the disk interior are made of 
silicate when $r<0.8$AU and ice-silicate for $r>0.8$AU.
 The fractional thickness of the water ice mantle,
$\Delta a/a$, is
set to 0.4 for both disk surface and disk interior.
Unlike C01, we use the refractive index of astronomical silicate 
(Draine \& Lee 1984; Draine 1985;
Loar \& Draine 1993; see also Weingartner \& Draine 2001).
We take optical constants of pure water ice from a NASA web site
(ftp://climate1.gsfc.nasa.gov/wiscombe).

The column density of the disk is $\Sigma_0 r_{AU}^{-3/2}$ with
$\Sigma_0=1000 g/cm^2$. 
Here $r_{AU}$ is distance measured in AU.
The disk is geometrically flared and 
the height of the disk surface is set to 4 times the disk scale height $h$.
The disk inner radius is $2R_*$ and the outer radius is $100$AU.
The central star has radius of $R_* = 2.5 R_{Sun}$ and temperature
of $T_* = 4000$K.
Temperature profile, flaring of disk, and
other details of the disk model
are described in C01.

\subsection{Radiative torque for large grains}

For most of the ISM problems,
dust grains are usually smaller than the wavelengths of interest.
However, this is no longer true in T Tauri disks because
we are dealing with grains as large as $\sim 1000 \mu m$.
To understand grain alignment in T Tauri disks we need to know
radiative torque for large grains.

In this study, we do not directly calculate radiative torque
for large grains.
Instead, we use a simple scaling relation to model
radiative torque for large grains.

\begin{figure}
\includegraphics[width=.48\textwidth]{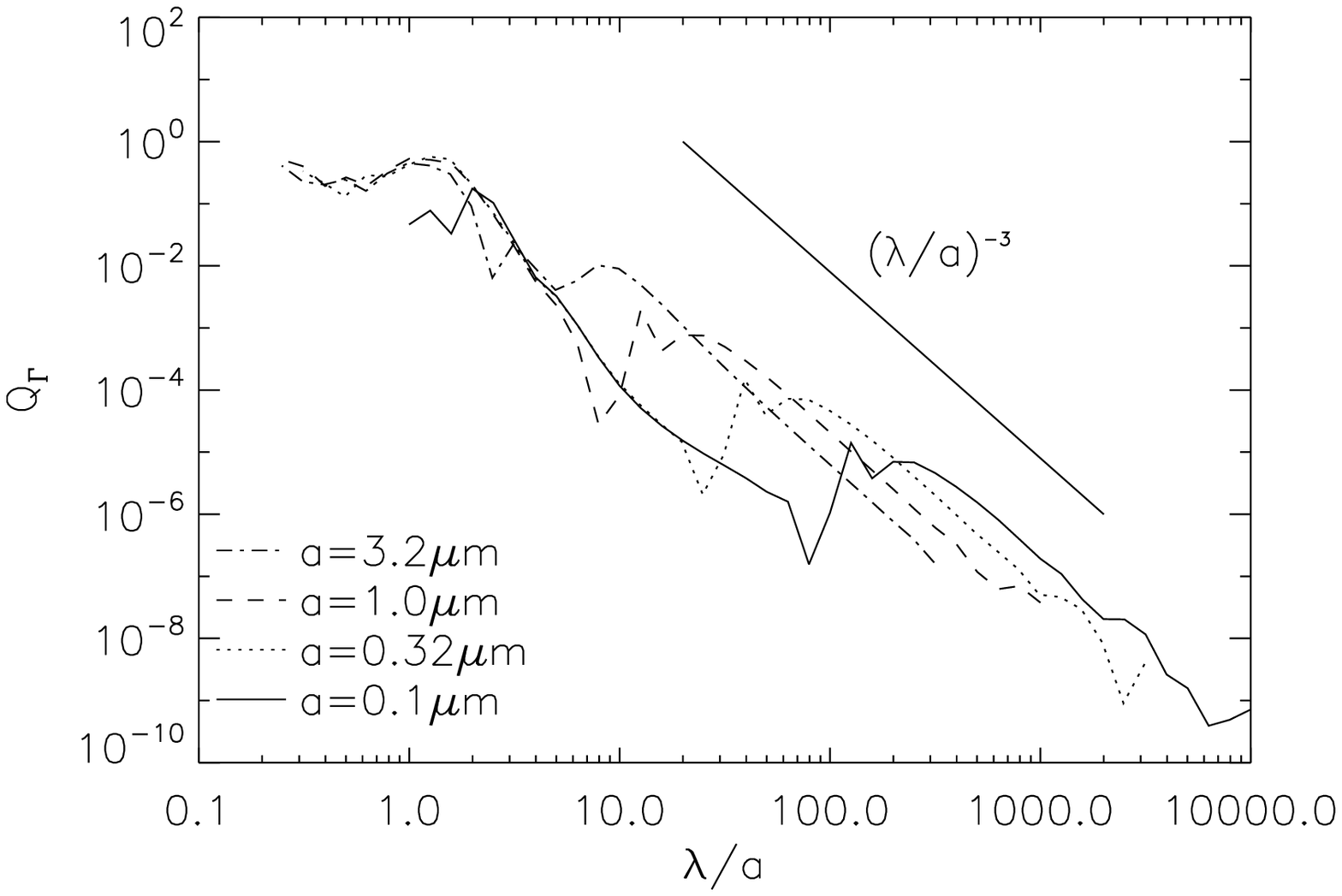}
\includegraphics[width=.48\textwidth]{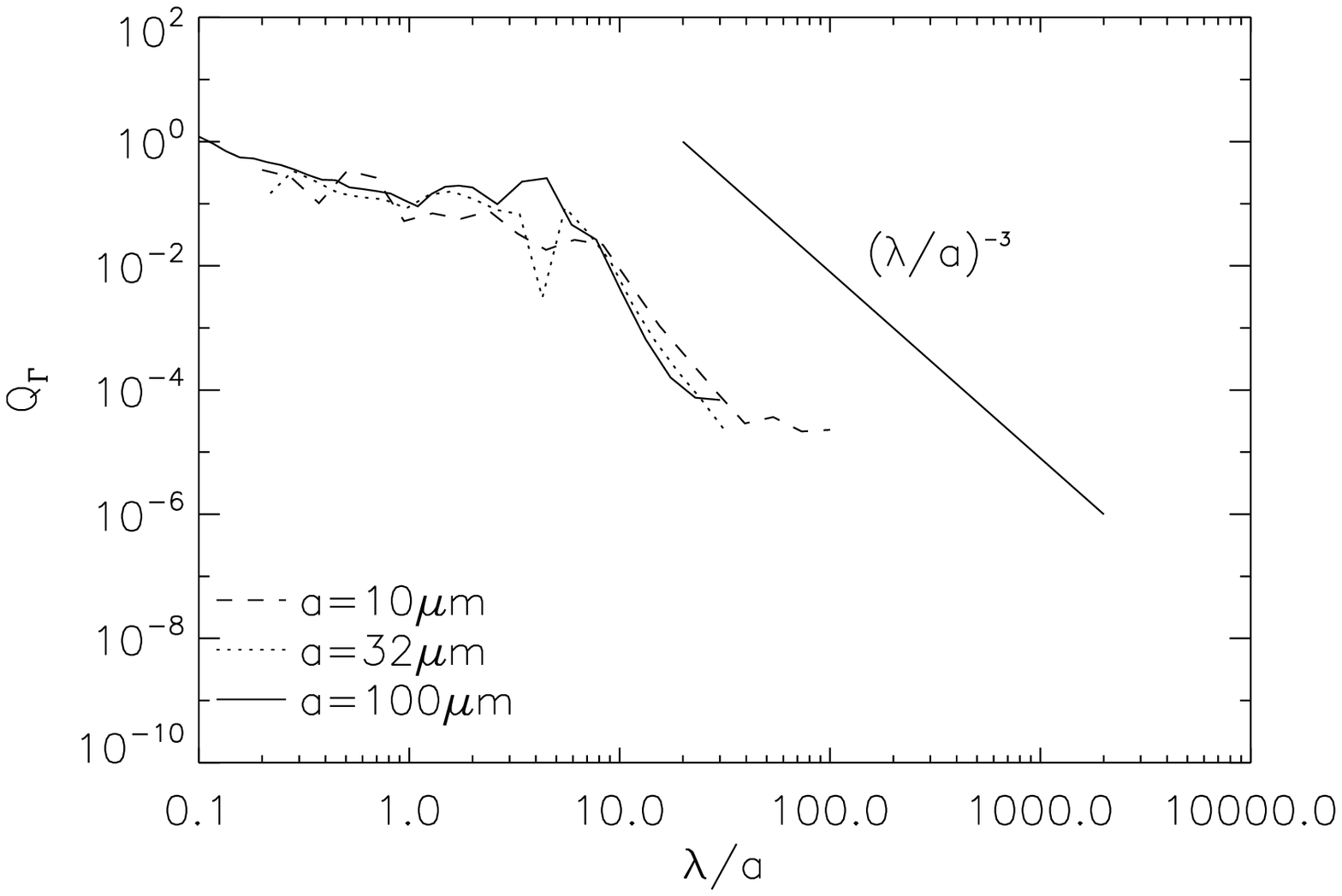}
\caption{
 Behavior of Torque.
Torque is $\sim O(1)$ when $\lambda \sim a$, 
where $a$ is the grain size.
Roughly speaking, torque $\propto (\lambda/a)^{-3}$.
{\it Left panel:} The results for small grains. 
    Data from CL05.
{\it Right panel:} The results for large grains.
    Data from Lazarian \& Hoang (2007).
\label{fig:tq}
}
\end{figure}

In CL05, we used the DDSCAT software package 
(Draine \& Flatau 1994; Draine \& Flatau (astro-ph/0409262); 
 Draine \& Weingartner 1996)
to calculate radiative torque on grain particles and showed 
the relation between $\lambda Q_{\Gamma}$ and $\lambda / a$
for grains with radii between $0.1 \mu m$ and $3.2 \mu m$.
Here $Q_{\Gamma}$ is the radiative torque efficiency.
Figure \ref{fig:tq}, obtained by reprocessing the earlier relation,
 shows that the radiative torque 
\begin{eqnarray}
 Q_\Gamma = \left\{ \begin{array}{ll}
                    \sim O(1) & \mbox{~~~ if $\lambda \sim a$} \\
                    \sim (\lambda/a)^{-3} & \mbox{~~~if $\lambda > a$,}
                    \end{array}
             \right.
\label{eq:tq}
\end{eqnarray}
where $a$ is the grain size and $\lambda$ the wavelength of the incident
radiation. Note that the 
radiative torque peaks near $\lambda \sim a$ and that
its value is of order unity there.
A more general study on this issue is provided in Lazarian \& Hoang (2007).
This allows us to assume that the relation holds true both for small and 
large grains.

\subsection{Rotation rate of dust grains by radiative torque}

To obtain grain rotational velocity one requires to calculate the balance between the excitation of rotation, driven by different processes and the
damping of rotational motions (see Draine \& Lazarian 1998). Below we 
shall calculate the rotational rates of a grain driven by radiative torques and subjected to gaseous damping. Note that, according to Lazarian \& Hoang (2007), the obtained rates will not be the actual rotational rates of a grain that is free to get aligned under the influence of the anisotropic radiation. In fact, most of the grains will be driven to low-$J$ 
attractor points and will rotate thermally or even sub-thermally. However, being driven by anisotropic radiation, grains get aligned and it was 
argued in Lazarian (2007) that the parameterization of the radiative 
torques in terms of the maximal achievable angular velocity is a valid 
one. Incidentally, this was also the parameterization that we used in
CL05 and in the subsequent paper by Bethell et al. (2007).

After some modifications, equation (67) in DW96
 reads
\begin{equation}
\left( \frac{ \omega_{rad} }{ \omega_T } \right)^2
 = 4.72 \times 10^9 \frac{ \alpha_1 }{ \delta^2 } \rho_3 a_{-5} 
  \left( \frac{ u_{rad} }{ n_H kT } \right)^2
  \left( \frac{ \lambda }{ \mu m } \right)^2  \\ \nonumber
   [ Q_{\Gamma} ]^2 
  \left( \frac{ \tau_{drag} }{ \tau_{drag,gas} } \right)^2,
\label{eq:ww}
\end{equation}
where $Q_{\Gamma} = \bf{Q}_{\Gamma} \cdot \hat{\bf{a}}_1$ and
$\hat{\bf{a}}_1$ is the principal axis with largest moment of
inertia, $n_H$ is the hydrogen number density, 
$u_{rad}$ is the energy density of the radiation field, 
$\delta \approx 2$, $\alpha_1 \approx 1.745$,
$\rho_3=rho/3g cm^{-3}$, $a_5=a/10^{-5}cm$,
and $\omega_T$ is the thermal angular frequency, which is the rate
at which the rotational kinetic energy of a grain is equal to $kT/2$.
The timescales 
$\tau_{drag,gas}$ and $\tau_{drag,em}$ are the damping time for gas drag and for electromagnetic emission, respectively, and
they satisfy the relation
$\tau_{drag}^{-1}=\tau_{drag,em}^{-1}+\tau_{drag,gas}^{-1}$  (see
Draine \& Weingartner (1996) for details).
As we discussed in the previous subsection, $Q_{\Gamma}$ is of
order of unity when $\lambda \sim a$ and declines as
$(\lambda/a)$ increases. From this observation, we can write
\begin{equation}
\left( \frac{ \omega_{rad} }{ \omega_T } \right)^2
\approx 
\left( \frac{ \omega_{rad} }{ \omega_T } \right)_{\lambda \sim a}^2
\left( \frac{ Q_{\Gamma, \lambda \sim a} }
            { Q_{\Gamma, \lambda}        } \right)^2
\approx 
\left( \frac{ \omega_{rad} }{ \omega_T } \right)_{\lambda \sim a}^2
\left( \frac{ \lambda  }{a} \right)^{-6}
\label{eq:tqlambda}
\end{equation}
for $\lambda > a$, where
\begin{equation}
\left( \frac{ \omega_{rad} }{ \omega_T } \right)_{\lambda \sim a}^2
 \approx 4.72 \times 10^9 \frac{ \alpha_1 }{ \delta^2 } \rho_3 a_{-5} 
  \left( \frac{ u_{rad} }{ n_H kT } \right)^2
  \left( \frac{ \lambda }{ \mu m } \right)^2  \\ \nonumber
  \left( \frac{ \tau_{drag} }{ \tau_{drag,gas} } \right)^2,
  \label{eq:tqla}
\end{equation}

\subsection{Minimum and maximum aligned grain size}
As in DW96, we assume that
grains are aligned when  $(\omega_{rad}/\omega_T)^2 > 10$.
Suppose that a monochromatic radiation field illuminates 
dust grains and that $(\omega_{rad}/\omega_T)^2 > 10$ for $\lambda \sim a$.
According to Eq.~(\ref{eq:ww}), the ratio $(\omega_{rad}/\omega_T)^2$
decrease as $a$ decreases and
the ratio will drop below $\sim 10$. 
Applying Eq.~(\ref{eq:tqlambda}) we can easily find the minimum aligned
grains size:
\begin{equation}
   a_{lower} \sim \lambda \left( \frac{10}
                      {(\omega_{rad}/\omega_T)_{\lambda \sim a}^2}
                         \right)^{1/6}
  \label{eq:lower}
\end{equation}

Then, how can we find the maximum aligned grain size?
In other words, are all grains with $a > \lambda$ aligned?
In order to answer this question, we need to consider
the behavior of torque $Q_{\Gamma}$ for the limit $ a \gg \lambda$.
Note that, when $\lambda \sim a$, $Q_{\Gamma} \sim O(1)$.
Then what happens when $ a \gg \lambda$?
In principle, we can calculate $Q_{\Gamma}$ using numerical simulations.
However, this is still infeasible because enormous computational power
is required. Therefore we can only conjecture what will happen
for $ a \gg \lambda$.

If one adopts the reasoning for the origin of RT in Dolginov \& Mytrophanov (1976)
one might expect that $Q_{\Gamma}$ drops due to incoherent 
contributions. Theoretical considerations in Lazarian \& Hoang (2007) show that this
may not be true. While to be conservative we adopt a rule of thumb is that $Q_{\Gamma}$ may begin to decline
when $a \geq 10 \lambda$, we will see below that this assumption does not alter
our results in any appreciable way\footnote{
Note that here $\lambda$ is the wavelength at which
radiation is strongest. For most cases, 
$\lambda = \lambda_{max,Wien}\propto 1/T$ when the radiation
field is a black body radiation, where $\lambda_{max,Wien}$ is the
$\lambda_{max}$ from Wien's law.}.

We also need to consider many different time-scales: precession time-scale, 
time-scale for the alignment of angular momentum and $\hat{\bf{a}}_1$,
etc. However, fortunately, the exact knowledge on
the maximum aligned grains size is not so important
for our current study (see next section).

\subsection{Grain alignment in disks}
We use Eq.~(\ref{eq:tq}), instead of the DDSCAT software package, 
to obtain radiative torque ($Q_{\Gamma}$) on grain particles in the T tauri disk
described in \S\ref{sect:diskmodel}.
We take a conservative value of $Q_{\Gamma}$ at $\lambda \sim a$:
 $Q_{\Gamma}\sim 0.1$ at $\lambda \sim a$.
Apart from $Q_{\Gamma}$, we also need to know $u_{rad}$ and 
$n_H$ to get the $(\omega_{rad}/\omega_T)_{\lambda \sim a}^2$ ratio
(see Eq.~(\ref{eq:tqla})). 
We directly calculate $u_{rad}$ and 
$n_H$ using the disk model in C01.
We assume that $\tau_{drag}\sim \tau_{drag,gas}$.

In the disk interior, there are two kinds of radiation fields:
one from the surface layer and the other from the disk interior itself.
We assume that both radiation fluxes direct only along the disk
vertical axis (i.e. ``z'' axis).
As in CG97, we assume that
half of the stellar radiation flux that reaches the disk surface 
enters the disk interior. The radiation flux from the star
is $\sim (\alpha/2)(R_*/r)^2 \sigma_B T_*^4$, where
$\alpha$ is the grazing angle at which the starlight strikes the
disk and $R_*$ is the stellar radius,
$T_*$ is the stellar temperature, and
$\sigma_B$ is the Stephan-Boltzmann constant (CG97).
We assume that the 
radiation flux from the surface layer has a narrow
spectrum around the wavelength of $\sim 3000/T_{ds} ~\mu m$,
where $T_{ds}$ is the grain temperature in the surface
layer (see C01 for a graph for $T_{ds}$).
The magnitude of the flux from the surface
at a given height is less than
the half of the incident stellar flux because the flux
from the surface is attenuated by
dust absorption in disk interior.
We also need to consider that there are two surface layers - one 
above the mid-plane and the other below it.
The flux from disk interior is also treated as a monochromatic wave 
with  $\lambda \sim 3000/T_i$ $\mu m$.
Note that flux from the interior has longer wavelengths
because the disk interior is cooler (see C01 for a 
graph for the disk interior $T_i$).
{}From the two fluxes, we can calculate anisotropic component
of the radiation energy density.
We use the fact that the disk surface density is
given by $\Sigma=1000 r_{AU}^{-3/2} cm^{-2}$ and that
the height of the disk $H(r)$ can be calculated from 
\begin{equation}
  H/r \sim 4 \sqrt{ T_i/T_c }\sqrt{r/R_*},
\end{equation}
where $T_c=3\times 10^{-24}GM_*/kR_*$ in cgs units.
Here $k$ is the Boltzmann constant.

\begin{figure}[h!t]
\includegraphics[width=.48\textwidth]{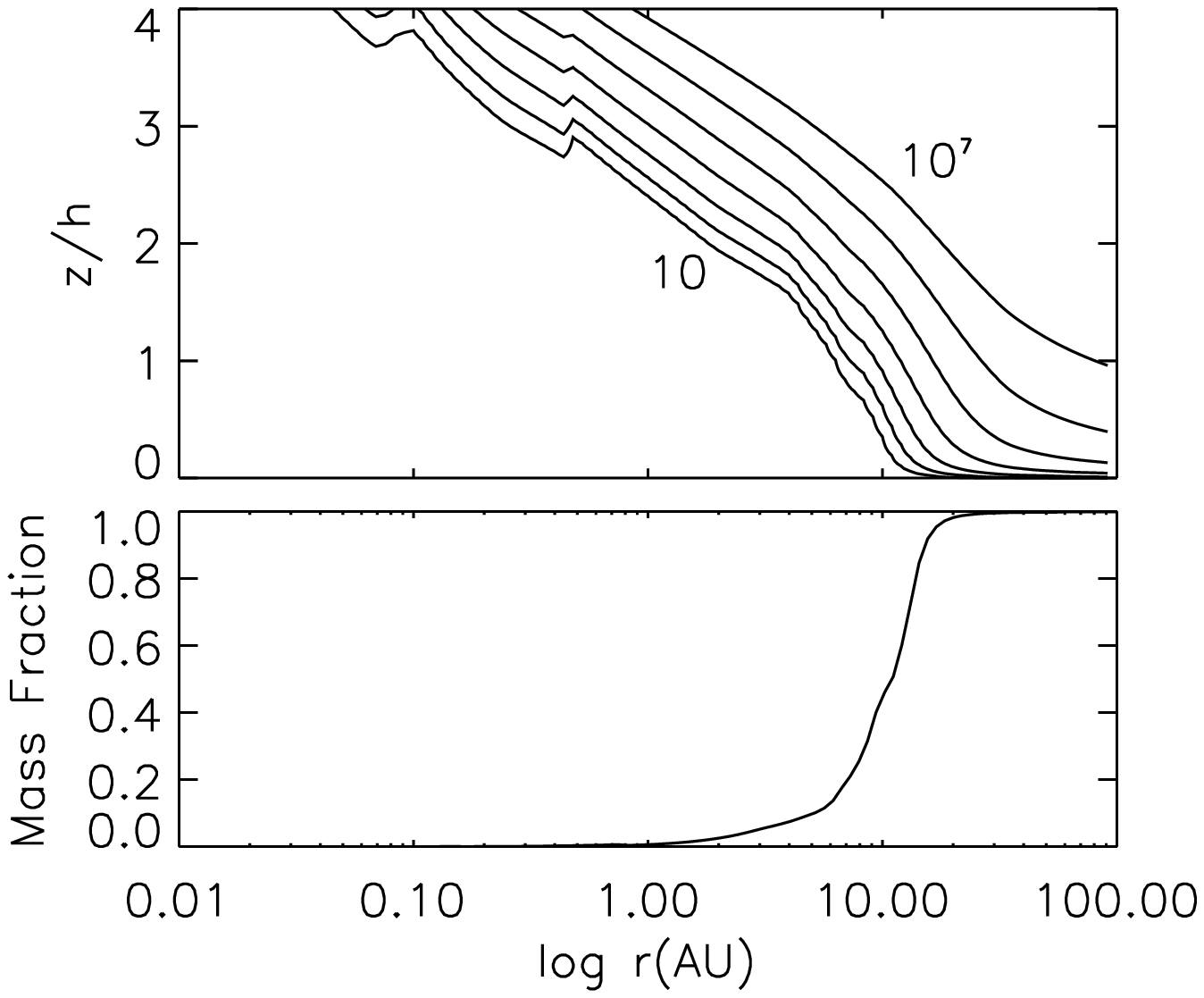}
\includegraphics[width=.48\textwidth]{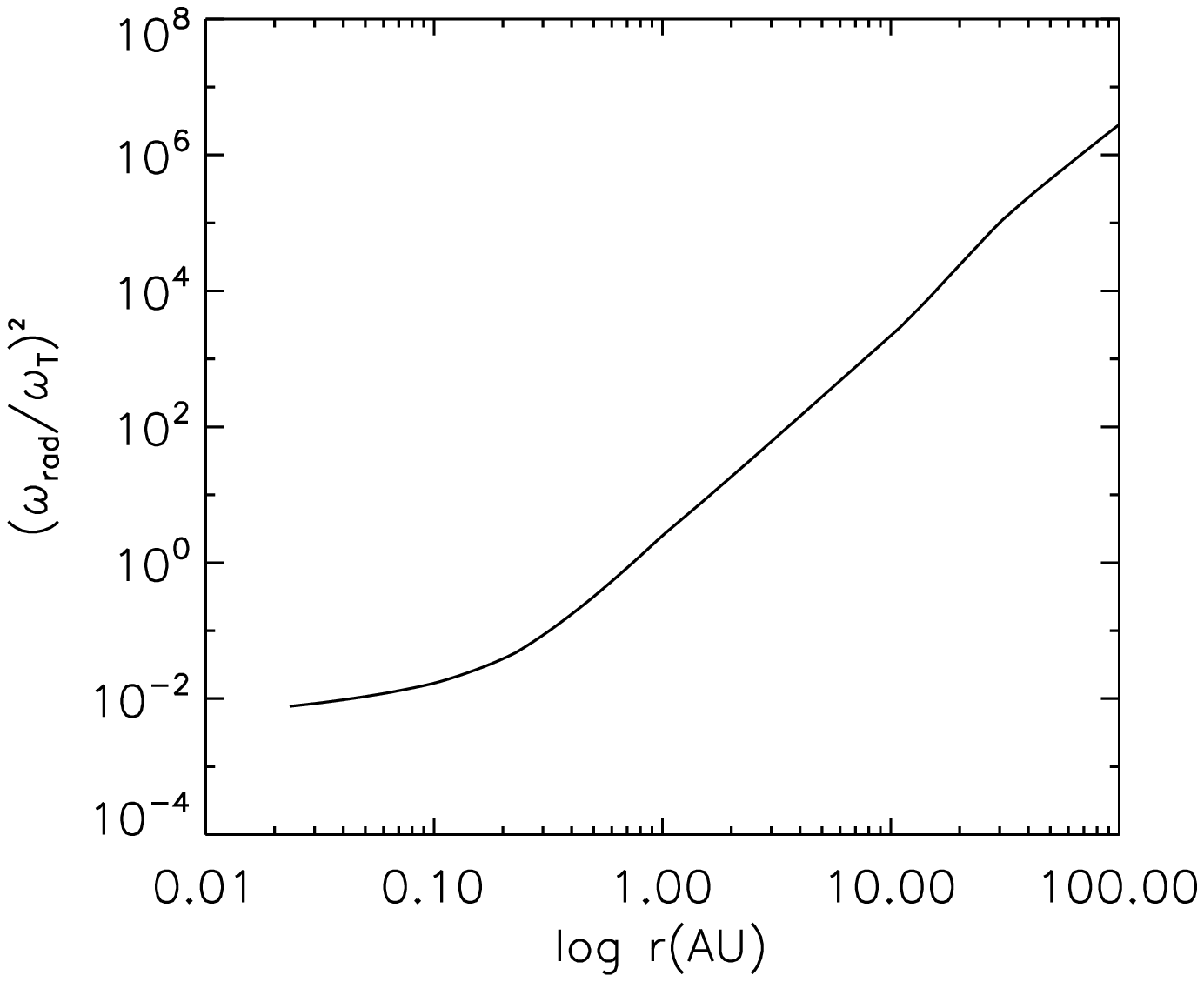}
\caption{Grain alignment in disk interior.
{\it Upper panel}: Contours show
the ratio $(\omega_{rad}/\omega_T)_{\lambda \sim a}^2$. Note that
the disk vertical height is shown in units of the disk scale height $h$.
We assume that grains are aligned when $(\omega_{rad}/\omega_T)^2> 10$.
{\it Lower panel}: Fraction of grains that satisfy 
$(\omega_{rad}/\omega_T)_{\lambda \sim a}^2 > 10$
 as a function of disk radius.
After $r> 10$AU, almost 100\% of grains satisfy the inequality.
\label{fig:align_int}
}
\caption{Grain alignment in surface layer. 
The ratio 
$(\omega_{rad}/\omega_T)_{\lambda \sim a}^2$
exceeds 10 when $r \geq 1$AU, which means that
some grains in the surface layer are aligned when  $r \geq 1$AU.
\label{fig:align_sur}
}
\end{figure}

We assume that grains are aligned when the ratio 
$(\omega_{rad}/\omega_T)_{\lambda \sim a}^2$ exceeds 10, which is overly conservative
according to a recent study in Hoang \& Lazarian (2007). 
Figure \ref{fig:align_int} shows that, at large $r$, large grains 
are aligned even deep inside the interior. 
Here we take $\lambda$ as the $\lambda_{max,Wien}$ of the local
 blackbody radiation field
at $r$.
On the other hand, at small $r$, only grains near the
disk surface are aligned.
This is because gas density is low and, therefore, the gas drag
is smaller near the surface layer.
The lower panel of Figure \ref{fig:align_int} shows that
almost all grains are aligned when $r> 10$AU.
Therefore, we expect strong polarized emission from
outer part of the disk.

In the surface layer,  grains are aligned by the
star-light. Note that the radiation field
scales as $r^{-2}$.
Since the column density scales as $r^{-3/2}$ and
the disk height is an increasing function of $r$, 
the density in disk surface will drop faster than $r^{-3/2}$.
The gas temperature in the surface layer
roughly scales as $r^{1/2}$ (see CG97 and C01).
Therefore we expect that  the ratio $\omega_{rad}/\omega_T$
is an increasing function of $r$ (see Eq.~(\ref{eq:ww})).
This implies that 
grains at large $r$ are aligned.
Grains near the central star cannot be aligned due to
high gas density near the star.
Indeed Figure \ref{fig:align_sur} shows the ratio 
$(\omega_{rad}/\omega_T)_{\lambda \sim a}^2$
exceeds 10 when $r \geq 1$AU, which means that
grains in the surface layer are aligned when  $r \geq 1$AU.
Note that the radiation from the central star has a 
$\lambda_{max,Wien}$ at $\sim 750 nm$.
We expect that polarized emission from the surface layer
is also originated from outer part of the disk.

\section{Theoretical Estimates of degree of polarization}

In this section, we estimate the degree
of polarization of emitted radiation in IR wavelengths. 
We need only the
grain size distribution for the calculation in this section. 
We do not use a detailed disk model.
The discussion in this section is applicable for
any system with large grains.

\begin{figure*}[h!t]
\includegraphics[width=.48\textwidth]{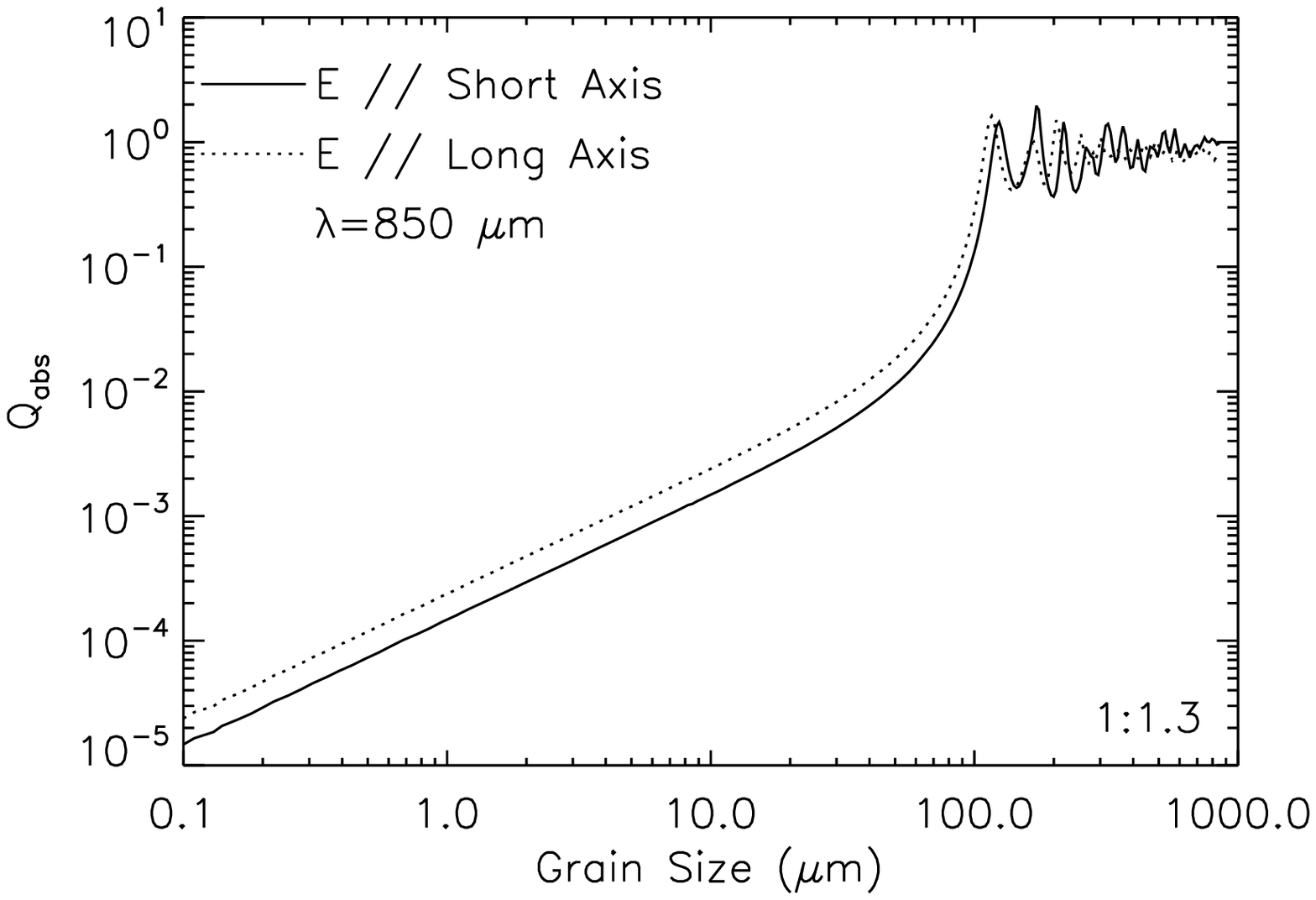}
\includegraphics[width=.48\textwidth]{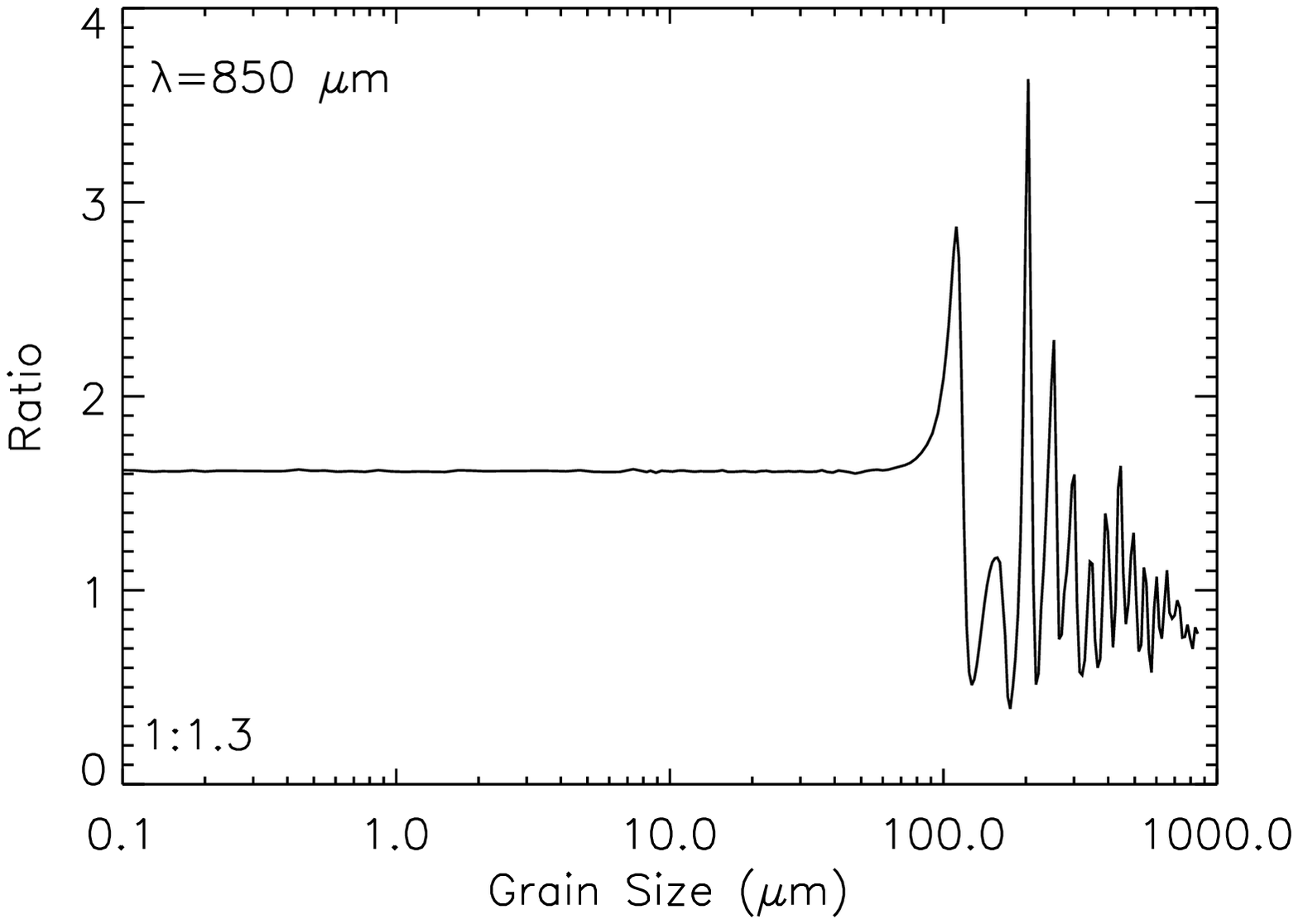}
\caption{Relation between the absorption cross-section 
   (=$Q_{abs}$ times the geometric cross-section) and grain size.
The wavelength at which we observe is 850 $\mu m$.
The grains are oblate spheroid made of silicate
with the symmetry axis $\hat{\bf{a}}_1$ parallel
to y-axis. Axis ratio is 1:1.3.
Radiation is propagating along x-axis.
{\it Left panel}:
When grains are smaller than $\sim 850/2\pi$, then
we can observe polarization because two cross-sections are
different.
When grains are larger than $\sim 850/2\pi$, then
we cannot observe polarization because  two cross-sections are
similar.
{\it Right panel}: When the axis ratio is 1:1.3, the ratio of $Q_{abs}$ is around
    $\sim 1.6$ when the grain is small compared with the wavelength 
(i.e. $2\pi a /\lambda <1$)  
    and $\sim 1$ for geometrical optics regime (i.e. $2\pi a/\lambda > 1$).
\label{fig:cratio}
}
\end{figure*}
\begin{figure*}[h!t]
\includegraphics[width=.48\textwidth]{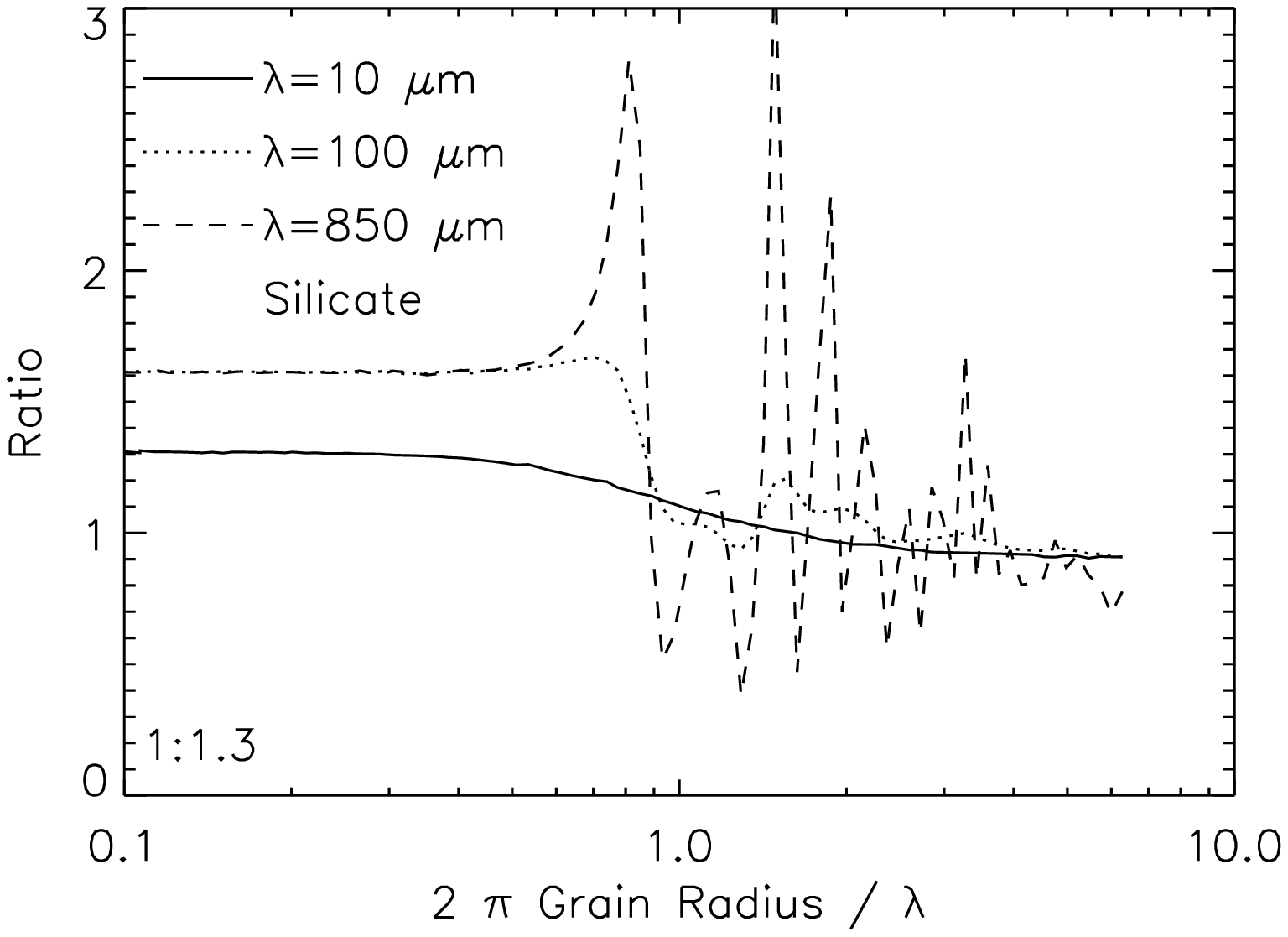}
\includegraphics[width=.48\textwidth]{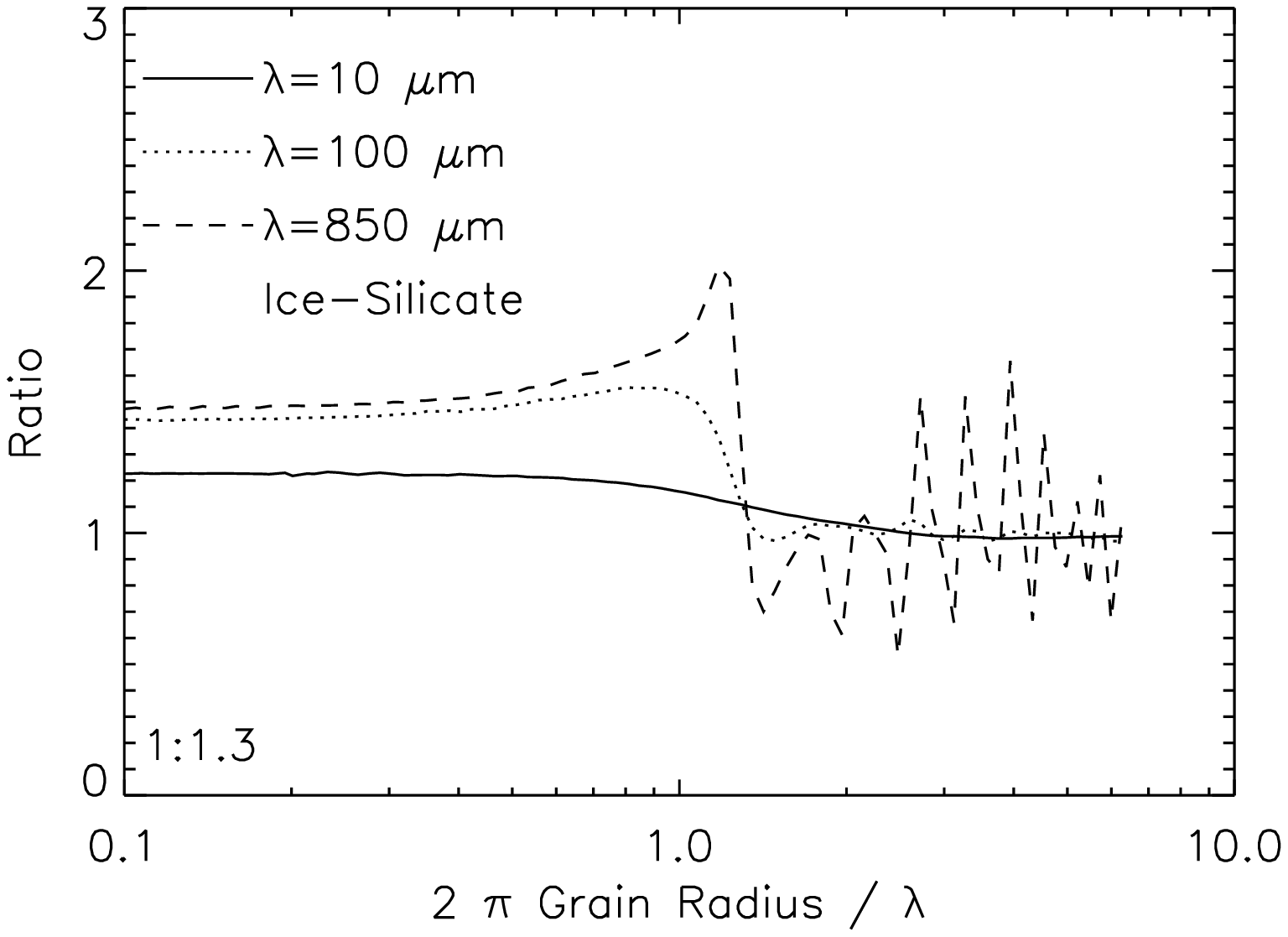}
\caption{The ratio $Q_{abs,\perp}/Q_{abs,\|}$ versus wavelength.
{\it Left panel}: Silicate grains.
For $\lambda \geq 100 \mu m$, the ratio is $\sim 1.6$
when the size parameter is less than 1 (i.e. $2\pi a/\lambda < 1$).
However, for shorter wavelengths, the ratio for $2\pi a/\lambda < 1$ drops:
it is $\sim 1.3$ for $\lambda=10\mu m$.
{\it Right panel}: Ice-Silicate grains (i.e. grains with silicate core and
water ice mantle).
\label{fig:moreratio}
}
\end{figure*}

Suppose that all grains are perfectly aligned.
Then, what will be the observed degree of polarization?
Of course, we do not observe 100\% polarization.
The size parameter $ 2\pi a/\lambda$ plays an important role here.
There are three important points that determine the degree of polarization
of emitted radiation. 

First, when the grain is small compared with the wavelength 
(i.e. when $2\pi a /\lambda < 1$), 
grain's intrinsic shape gives a limit.
Suppose that grains are oblate spheroid, that
the symmetry axis of the grain is parallel 
to y-axis, and that
radiation is propagating along x-axis.
When we send two linearly polarized radiation fields with electric field 
parallel to and perpendicular to the grain's symmetry axis, respectively, 
the radiation fields experience different cross-sections:
the radiation with ${\bf E} \parallel \hat{\bf a}_1$ sees smaller cross-section.
Here ${\bf E}$ is the electric field of the radiation field and $\hat{\bf a}_1$
is the grain's symmetry axis.
As a result, we observe polarization because two cross-sections are
different. So far we have dealt with polarization by absorption.
Polarization by emission is caused by the exactly same fact 
that two cross-sections are different.
Since $Q_{abs}=Q_{em}$, where $Q_{em}$ is the grain emissivity,
grain emits more radiation with ${\bf E} \perp \hat{\bf a}_1$ than
one with  ${\bf E} \parallel \hat{\bf a}_1$
The degree of polarization of emitted light is
$(Q_{abs,\perp}-Q_{abs,\|})/(Q_{abs,\perp}+Q_{abs,\|})$,
where $\|$ and $\perp$ refer to directions parallel and perpendicular
to the grain's symmetry axis $\hat{\bf a}_1$.

Figure \ref{fig:cratio}, obtained from the DDSCAT package,
shows this effect clearly.
The the radiation fields have $\lambda = 850\mu m$ and 
 the grains' long to short axis ratio is 1.3:1.
The ratio
of two cross-sections is around $1.6$ for $ 2\pi a /\lambda < 1$.
When we observe emission from those grains,
the degree of polarization can be as large as (1.6-1)/(1.6+1)=22\%.
If we assume that grains' long to short axis ratio is 1.5:1, 
the ratio of the cross-sections is
2.1:1 and the resulting degree of polarization for emission is
as large as (2.1-1)/(2.1+1)=35\%.
The ratio of the cross-sections varies as wavelength of radiation varies.
The ratio seems to be fairly constant for $\lambda > 100\mu m$.
However, for shorter wavelengths, the ratio decreases (Figure \ref{fig:moreratio}).
The shape of grains in protostellar disk is uncertain 
(see, for example, Hildebrand \& Dragovan 1995 for the general ISM
cloud cases).

Second,  when $ 2\pi a /\lambda > 1$
(i.e. in the geometrical optics regime), we do not observe
polarization.  Figure \ref{fig:cratio} clearly shows that
the ratio of cross-sections becomes very close to 1 when
$ 2\pi a /\lambda > 1$.
This means that 
the usual argument about polarization by absorption or emission
works only when the grain size is small compared with the wavelength: 
$2\pi a /\lambda < 1$.
That is, in the small size parameter case, 
when radiation meets an elongated grain, 
it recognizes the elongated
shape and interacts differently depending on the direction of
the electric field of the radiation.
However, in the geometrical optics regime (i.e. when $2\pi a/\lambda > 1$), 
radiation does not recognize that grains are elongated.
We can easily understand this fact when we consider an elongated
macroscopic object: Cross-sections are same regardless of the
electric field directions for visible light.
This second point is somewhat tricky: Even in the case grains are 
actually ``aligned", we do not ``observe" polarization when
the grains are large compared with the wavelength.

This observation has an important consequence.
When we calculate polarization, large grains (i.e. grains with
$a > \lambda/2 \pi$) do not contribute to polarization even when
they are aligned.
This fact reduces the degree of polarization significantly
in protostellar disks.
For example, suppose that we has grains as large as $1000 \mu m$
in a disk.
If we observe emission from the disk at $\lambda=850\mu m$,
grains with $a> 850/2\pi \sim 100 \mu m$ do not contribute
polarization although they dominate extinction when 
the grain size distribution is a power-law ($dN\propto a^{-q} da$)
with $q< 3$.

Thirds, if the medium is optically thick,
the degree of polarization reduces.
The intensity of radiation from a uniform slab
is $S_{\nu} (1-exp(-\tau))$, where $S_{\nu}$ is the source function
and $\tau$ is the optical depth.
We can observe polarization when $\tau$, is different
for parallel and perpendicular (to grain symmetry axis or
generally any spatial direction) directions.
However, in the optically thick limit, the intensity becomes equal to $S_{\nu}$.
Therefore, difference in optical depth does not produce
observable level of polarization when the slab is opaque.

\begin{figure*}[h!t]
\includegraphics[width=.48\textwidth]{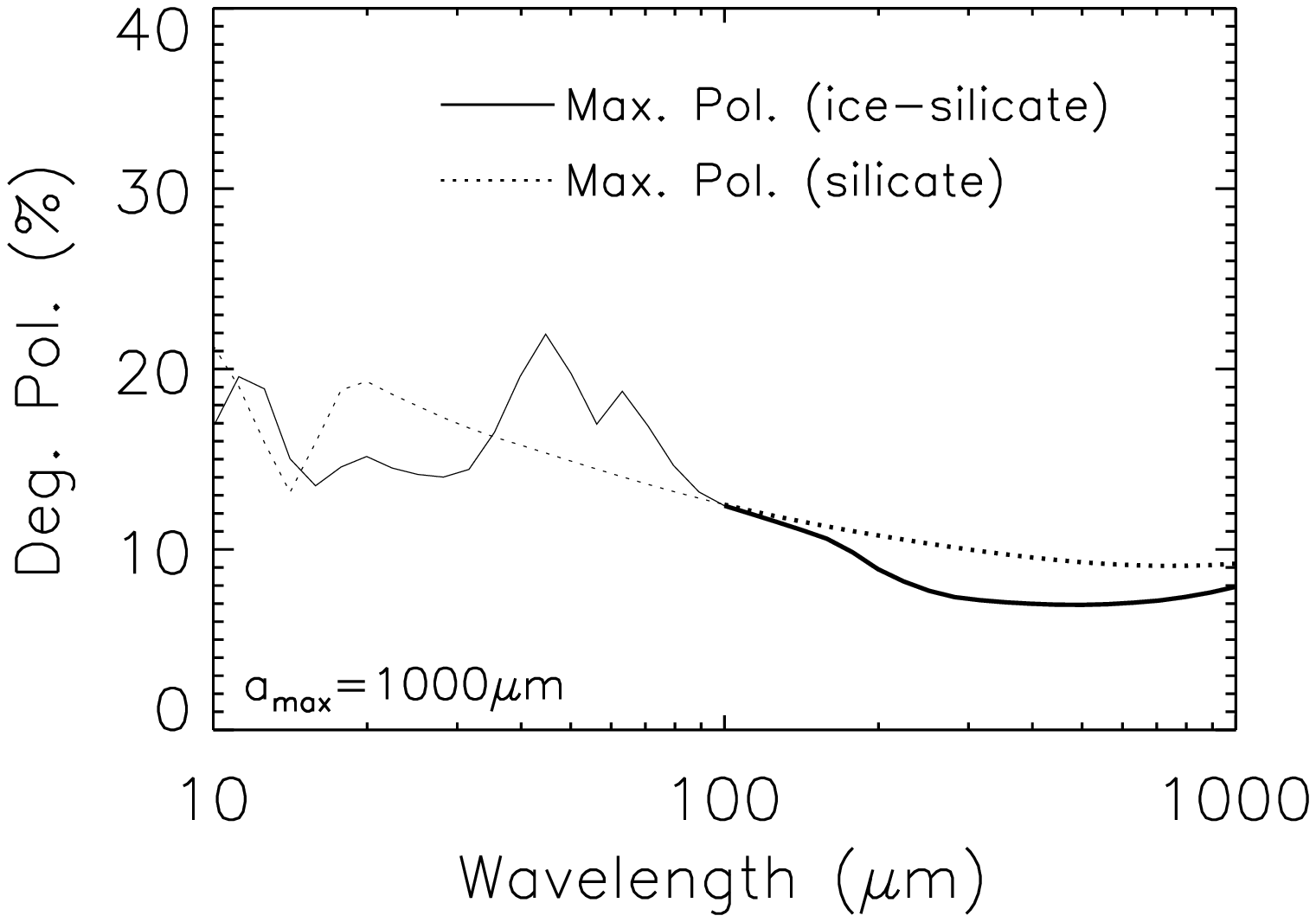}
\includegraphics[width=.48\textwidth]{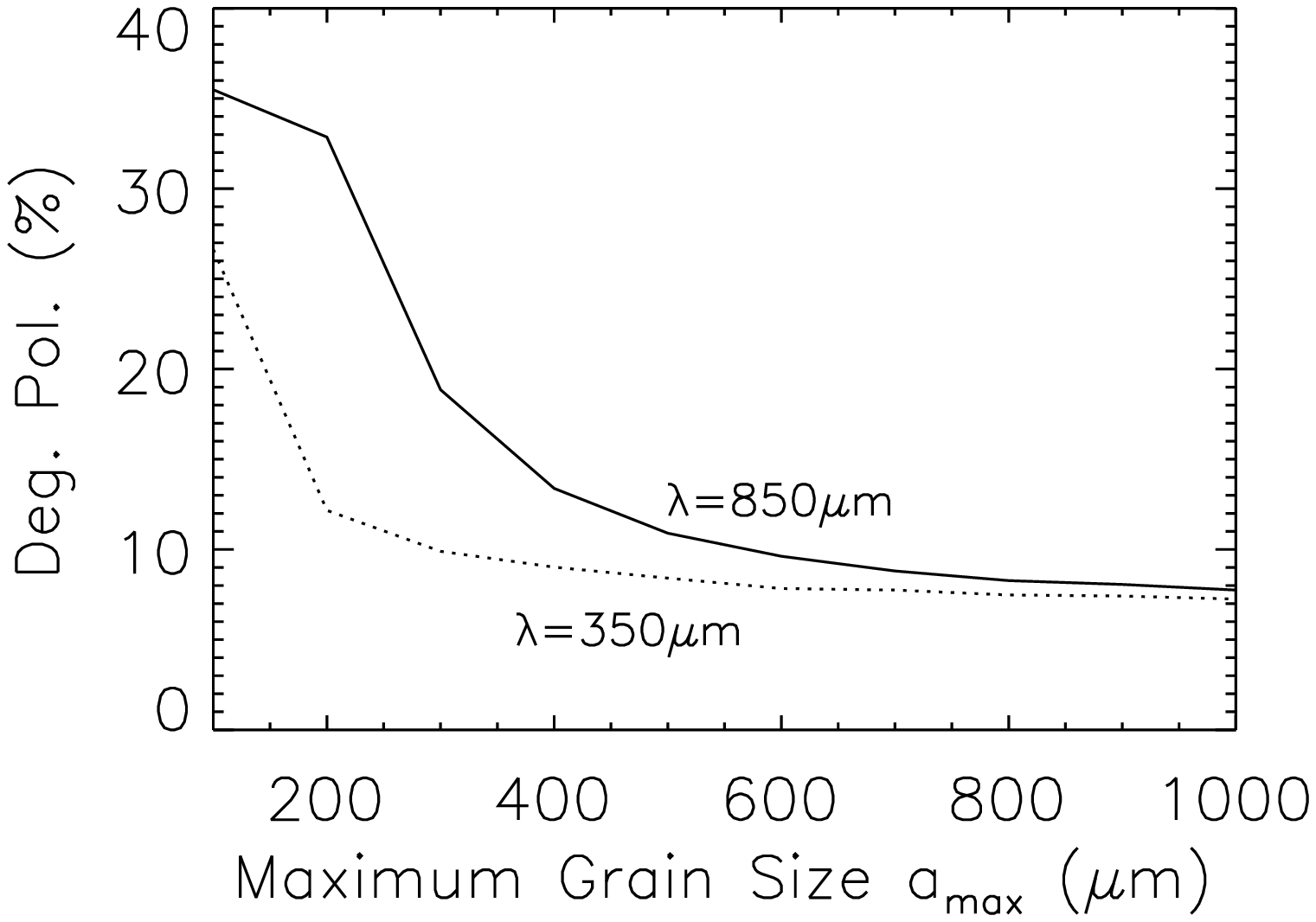}
\caption{Expected maximum degree of polarization for disk interior.
{\it Left panel}: $a_{max}=1000\mu m$. We assume grains are
    oblate spheroid with axis ratio of 1.5:1.
{\it Right panel}: The degree of polarization is very sensitive to the
    maximum grain size, $a_{max}$. Results are for ice-silicate.
   When  $a_{max}$ is smaller, more
grains are in the geometrical optics regime 
and, hence, the degree
of polarization rises for a given observing wavelength.
\label{fig:max}
}
\end{figure*}

For the {\it disk interior}, let us consider only the first and the second
points mentioned above. That is, let us assume that
the grains are perfectly aligned and the disk is optically thin.
We can estimate the degree of polarization $p$ for
the disk interior by integrating the following:
\begin{equation}
    p(\lambda) = \frac{  \int_{a_{min}}^{\lambda/2\pi}
        \left( Q_{abs,\perp}(a)- Q_{abs,\|}(a) \right) 
       a^2 N(a)da }{
                  \int_{a_{min}}^{a_{max}} 
        \left( Q_{abs,\perp}(a)+Q_{abs,\|}(a) \right) a^2 N(a)da   },
\end{equation}
where $Q_{...}$ is grain absorption efficiency, or grain emissivity, 
and $a_{min}=0.01 \mu m$.
In left panel of Figure \ref{fig:max}, we use $a_{max}=1000 \mu m$.
We use $dN\propto a^{-q} da$ with $q=3.5$ and assume that
grains are oblate spheroid with long to short axis ratio of
1.5:1.
It is not surprising that the degree of polarization rises
when we use smaller $a_{max}$: when $a_{max}$ is smaller, more
grains are in the geometrical optics regime.
Right panel of Figure \ref{fig:max} shows this effect.

Note that we do not use actual disk models here.
Actual numerical simulations using actual disk models 
will give smaller values because
not all grains are aligned and some part of the disk is optically thick.

Left panel of Figure \ref{fig:max} shows that
the maximum degree of polarization 
for $\lambda \le 100 \mu m$ is slightly
larger than that for $\lambda \ge 100 \mu m$.
However, in reality, we do not expect a significant degree of polarization
for $\lambda < 100 \mu m$. The reason is that
the entire disk becomes optically thick for $\lambda < 100 \mu m$.
The opacity per unit mass for $\lambda=100\mu m$ is
around $\sim O(0.1)$ (see Figure 3 in C01).
The outer most disk has column density of ~ 1g/cm$^2$.
Therefore, even the outer-most part of the
disk becomes optically thick when the wavelength drops below
$\sim 100\mu m$.

For the {\it disk surface layer}, 
the second point mentioned above is irrelevant
because grains are smaller in the surface layer: $a_{max}=1\mu m$.
The third point is also irrelevant because 
the surface layer is optically thin at far-infrared and submillimeter wavelengths.
Therefore, if all grains are perfectly aligned, the degree of polarization
of the emitted radiation
is determined only by the grain shapes (see the first point above).

\section{Estimates for Spectral Energy Distribution}

In this section, we calculate the degree of polarization of
emitted infrared radiation from a disk with structure and 
parameters described in
C01.
In this section, we assume that the disk in face-on.
The degree of polarization will be zero for a face-on disk when
magnetic field is perfectly azimuthal 
and the disk is cylindrically symmetric.
In this section, {\it we are concerned only with the
 absolute magnitude of the polarization}.

\subsection{Spectral energy distribution}

When $(\omega_{rad}/\omega_T)_{\lambda \sim a}^2$ is larger than $10$,
we find the minimum aligned size from Eq. (\ref{eq:lower}):
\begin{equation}
    a_{lower} \sim
\left( \frac{ 10 }{ (\omega_{rad} /\omega_T)_{\lambda \sim a}^2 }
\right)^{1/6} \lambda_{max,Wien},
\end{equation}
where $\lambda_{max,Wien}$ is the peak wavelength of the aligning radiation.
Grains smaller than $a_{lower}$ are not aligned.

When $(\omega_{rad}/\omega_T)_{\lambda \sim a}^2$ is larger than $10$,
we find the maximum size of grains that give rise to polarization from
\begin{equation}
  a_{upper} = min[ 10 \lambda_{max,Wien}, \lambda_{obs}/2\pi ],
\end{equation}
where $\lambda_{obs}$ the observing wavelength.
In most cases, $a_{upper} = \lambda_{obs}/2\pi $ because
$\lambda_{max,Wien}$ falls in far-infrared wavelengths.
Note again that the actual maximum aligned size can be larger 
than $a_{upper}$.

Now we know $a_{upper}$ and $a_{lower}$.
Note that $a_{upper}$ and $a_{lower}$ are functions of the distance
to the central star, $r$,
and the distance to the disk mid-plane, $z$.
We calculate the parallel (with respect to the local magnetic field) 
and perpendicular opacity
respectively:
\begin{eqnarray}
  \tau_{\|}(r,z) \propto \int_{a_{min}}^{a_{max}}
     Q_{abs}(\pi a^2) N(a) f_{\|} da  , \\
  \tau_{\perp}(r,z) \propto \int_{a_{min}}^{a_{max}}
     Q_{abs}(\pi a^2) N(a) f_{\perp} da,
\end{eqnarray}
where $N(a)da \propto a^{-3.5}$ (i.e. an MRN-type distribution) and
\begin{eqnarray}
 f_{\|} = \left\{ \begin{array}{ll}
             \sim 0.77 & \mbox{~(or $\sim 0.65$)~~ if $a_{lower}<a<a_{upper}$} \\
             1 & \mbox{~~~otherwise,}
                    \end{array}
             \right.
\end{eqnarray}
and 
\begin{eqnarray}
 f_{\perp} = \left\{ \begin{array}{ll}
             \sim 1.23 & \mbox{~(or $\sim 1.35$)~~ if $a_{lower}<a<a_{upper}$} \\
             1 & \mbox{~~~otherwise,}
                    \end{array}
             \right.
\end{eqnarray}
where we assume that the long to short axis ratio of the oblate spheroid
is 1.3:1 (or 1.5:1).

{}From this, we can calculate emission in parallel and
perpendicular directions:
\begin{eqnarray}
  L_{\lambda, \|} \propto  \lambda \int_{r_{min}}^{r_{max}} dr ~r
                        \int_{-4h}^{4h} dz 
                        \frac{ d\tau_{\lambda,\|} }{ dz } e^{-\tau_{\|}}
                        B_{\lambda}(T), \\
  L_{\lambda, \perp} \propto  \lambda \int_{r_{min}}^{r_{max}} dr ~r
                        \int_{-4h}^{4h} dz 
                        \frac{ d\tau_{\lambda,\perp} }{ dz } 
                        e^{-\tau_{\perp}}
                        B_{\lambda}(T),
\end{eqnarray}
where $\tau_{\|}$ and $\tau_{\|}$ measure optical depths from $z$ to
$4h$ along the axis perpendicular to the disk mid-plane (see 
CG97).
We use BHCOAT.f and BHMIE.f codes
in Bohren \& Huffman (1983) to calculate
grain emissivity $Q_{abs}$.

\begin{figure*}[h!t]
\includegraphics[width=.50\textwidth]{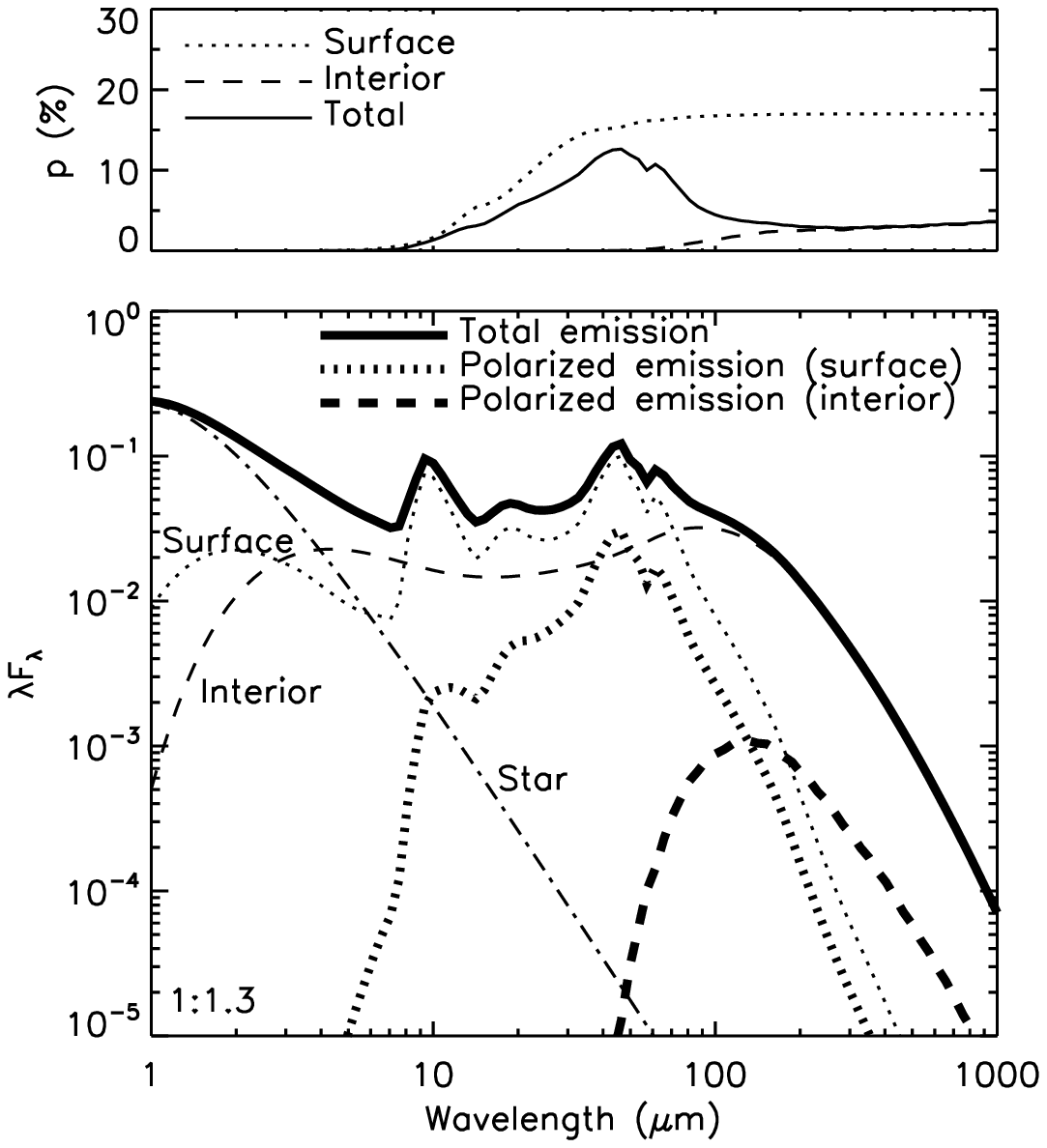}
\includegraphics[width=.50\textwidth]{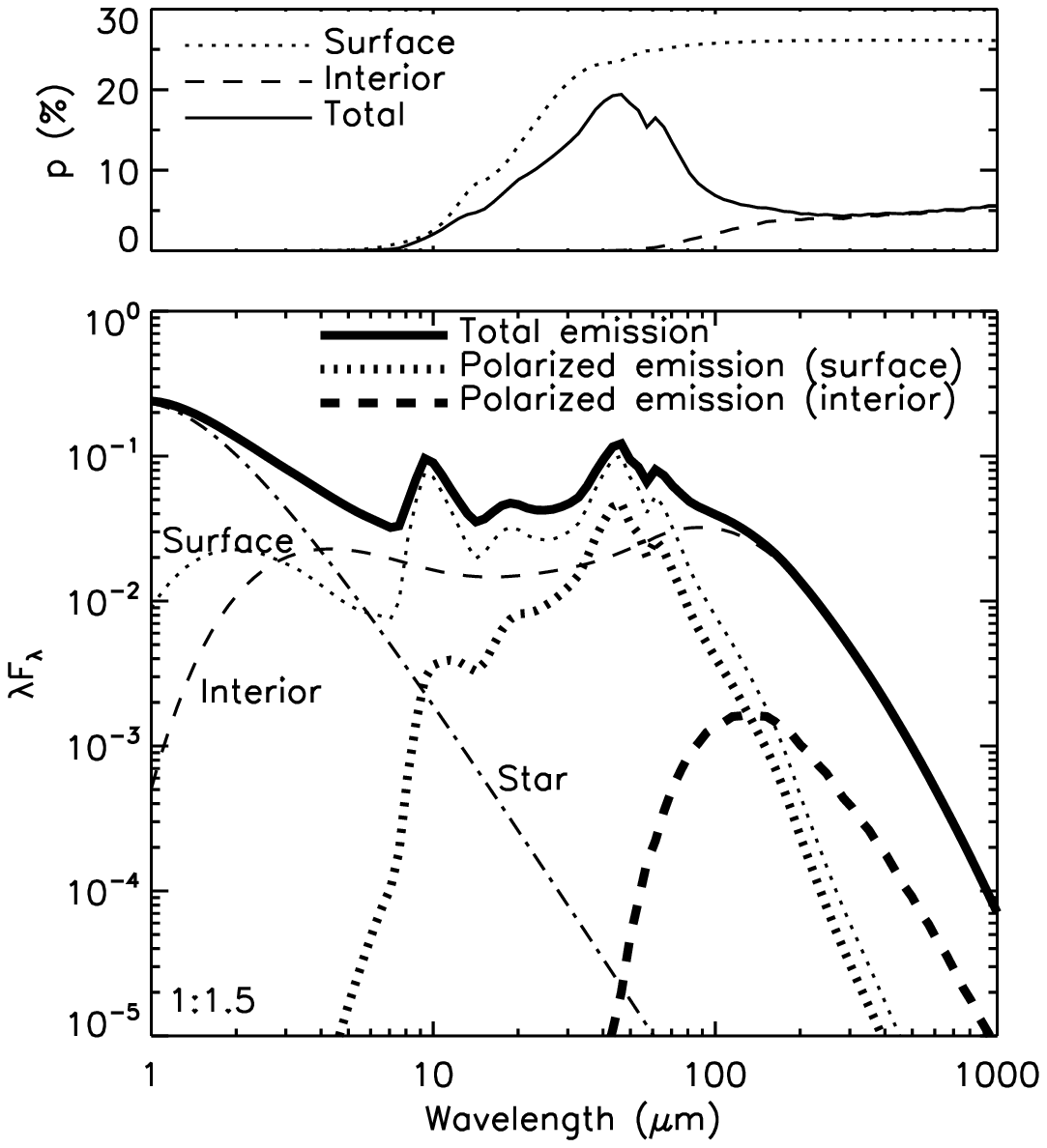}
\caption{Spectral energy distribution.
The vertical axis (i.e. $\lambda F_{\lambda}$)
is in arbitrary unit.
Thick solid line: total (i.e.~interior + surface) emission from disk.
Thin dotted line: total emission from disk surface.
Thick dotted line: polarized emission from disk surface.
Thin dashed line: total emission from disk interior.
Thick dashed line: polarized emission from disk interior.
Note that,  in these calculations of polarized emission, 
we ignored the direction
of polarization vectors and we only take the absolute value of them.
{\it Left panel:}
  Results for oblate spheroid grains with axis ratio of 1.3:1.
{\it Right panel:}
  Results for oblate spheroid grains with axis ratio of 1.5:1.
  The degree of polarization is larger than that in
  the left panel.
\label{fig:sed16}
}
\end{figure*}

The degree of polarization is 
\begin{equation}
   p(\lambda) = (L_{\lambda, \perp}-L_{\lambda, \|})/
                (L_{\lambda, \perp}+L_{\lambda, \|}).
\end{equation}
Figure \ref{fig:sed16} ({\it left panel} for 1.3:1 oblate spheroid and
{\it right panel} for 1.5:1 oblate spheroid) 
shows the results.
The degree of polarization can be as large as $\sim$5\% in FIR/sub-millimeter
wavelengths and $\sim$ 10\% in mid-IR regimes.
The polarized emission at FIR is dominated by the disk interior and
that at mid-IR is dominated by the disk surface layer.
Note again that, in these calculations, we ignored the direction
of polarization and we only take the absolute value of it.

\subsection{Radial energy distribution}

\begin{figure*}[h!t]
\includegraphics[width=.88\textwidth]{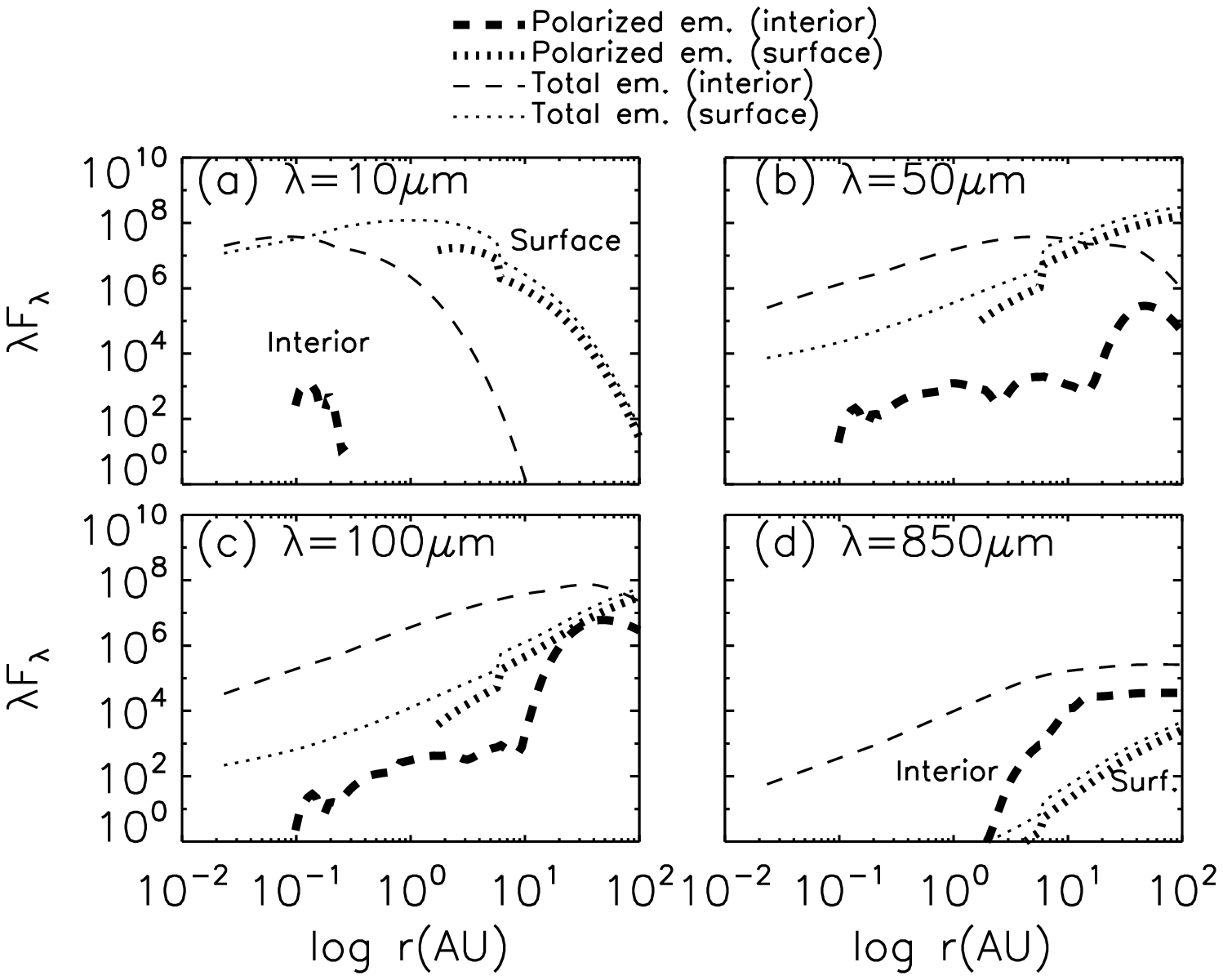}
\caption{Radial energy distribution.
(a) $\lambda=10 \mu m$. Inner part of the disk
   emits substantial amount of radiation. 
   But it emits negligible amount of polarized radiation.  
   Note that, when $r<1$AU, grains in the surface layer
   are not aligned and only negligible fraction of
   grains are aligned in the interior
   (see Figures \ref{fig:align_int} and \ref{fig:align_sur}).
(b) $\lambda=50 \mu m$.
(c) $\lambda=100 \mu m$.  
(d) $\lambda=850 \mu m$.
 The result for $\lambda=450 \mu m$ (not shown) is very similar to
 that for  $\lambda=850 \mu m$.
\label{fig:red}
}
\end{figure*}

Figure \ref{fig:red} shows radial distribution of emitted radiation.
For $\lambda=850 \mu m$, both radiations from the disk interior and
the surface layer are dominated by the outer part of the disk.
But, for $\lambda=10 \mu m$, the inner part of the disk
contributes significant portion of total emission and polarized emission.

\section{Effects of disk inclination}

In this section we calculate actual degree of polarization
that we can observe.
Chiang \& Goldreich (1999) calculated spectral energy
distribution (SED)
from inclined disks.
We follow a similar method to calculate the
the SED of
polarized emission. The SED
of disk interior
is the integral of the source function:
\begin{equation}
   L_{\lambda}^{int} \propto \lambda
    \int_{-r_{max}}^{r_{max}} dx
    \int_{-y(x)}^{y(x)} dy
    \int d\tau_{\lambda} B_{\lambda}(T_i) e^{ - \tau_{\lambda} },
\label{eq:sedang_int}
\end{equation}
where 
\begin{equation}
   y(x)=\sqrt{ r_{max}^2-x^2 } \cos\theta - H(r_{max}) \sin\theta,
\end{equation}
and $r_{max}$ is the outer disk radius, $H(r_{max})$ is the
height of disk at the disk outer radius,
$\theta$ is the angle between disk symmetry axis and the line 
of sight,
$T_i$ is the temperature of disk interior, $B_{\lambda}$
is the Planck function, and $\tau$ is the optical depth.
The integral $\int d\tau_{\lambda} ...$ is taken over the line of sight 
(see Chiang \& Goldreich 1999 for details).
The SED of the disk surface is obtained from
the integral
\begin{equation}
      L_{\lambda}^{surf} \propto \lambda
    \int_{-r_{max}}^{r_{max}} dx
    \int_{-y(x)}^{y(x)} dy
    \sum  B_{\lambda}(T_{ds}) 
         \left[ 1-exp\left(-\frac{ \alpha \epsilon_s}
                                 { |\hat{\bf n}\cdot \hat{\bf l}| }
                     \right)
         \right]
         exp( - \tau_{\lambda} ),
\label{eq:sedang_surf}
\end{equation}
where $\hat{\bf n}$  and $\hat{\bf l}$ are unit vectors 
normal to the surface and parallel to the line of sight, respectively,
$\epsilon_s$ is the Planck averaged dust emissivity at the surface.
The summation is performed whenever the line of sight intersects
the surface (see Chiang \& Goldreich 1999 for details).

In our calculations, we explicitly
take care of the fact that grain symmetric axis is changing
along a given line of sight.
We follow the description in Roberge \& Lazarian (1999) 
(see also Lee \& Draine, 1985) to 
calculate this effect and we obtain optical depths with respect to
the x and y directions in the above integrals 
(Eqs. (\ref{eq:sedang_int}) and (\ref{eq:sedang_surf}))\footnote
{Let the z-axis be the direction of the line of sight and 
the $y_0$ axis the direction along the projection of the magnetic field 
on to the plane of the sky. 
The $x_0$ axis is perpendicular to both axes.
Then, the cross sections $C_{x0}$ and $C_{y0}$ are
\begin{eqnarray}
C_{x0} = C_{avg} + \frac{1}{3} R (C_{\perp}-C_{\|})(1-3 \cos^2 \zeta)
, \nonumber \\
C_{y0} = C_{avg} + \frac{1}{3} R (C_{\perp}-C_{\|}), \nonumber
\end{eqnarray}
where $C_{avg}=(2C_{\perp}+C_{\|})/3$, $C_{\perp}$ and $C_{\|}$ are
cross sections with respect to the magnetic field, $\zeta$ is the angle
between the magnetic field and the plane of the sky. We assume the
Rayleigh reduction factor, $R$, is 1.
For the axes $x$ and $y$ that also lie in the plane of the sky,
\begin{eqnarray}
C_{x} \approx C_{x0} \cos^2 \theta + C_{y0} \sin^2 \theta, \nonumber \\
C_{y} \approx C_{x0} \sin^2 \theta + C_{y0} \cos^2 \theta, \nonumber
\end{eqnarray}
where $\theta$ is the angle between the $x$ and $x_0$ axis.
}.
After calculating the optical depths, we calculate 
$L_{\lambda, x}$ and $L_{\lambda, y}$.
We obtain the luminosity for two more directions, which have 45 degrees
with respect to x- and y-directions. 
We calculate
the the direction and  degree of polarization based on the luminosity for
these 4 directions. 

\begin{figure*}[h!t]
\includegraphics[width=.88\textwidth]{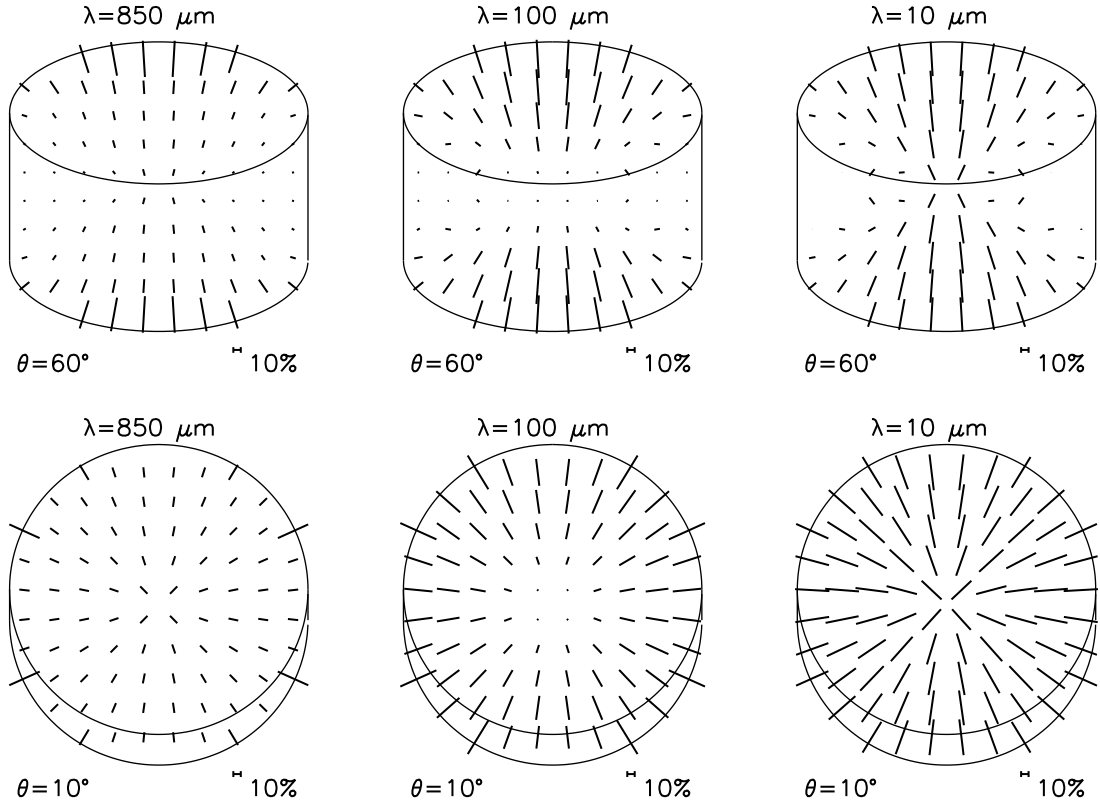}
\caption{
  Simulated observations. Degree of polarization is calculated for
the total radiation (i.e. interior + surface) from the disk.
 The disk inclination angle $\theta$ is the angle between
 disk symmetry axis and the line of sight.
\label{fig:visual}
}
\end{figure*}

\begin{figure*}[h!t]
\includegraphics[width=.88\textwidth]{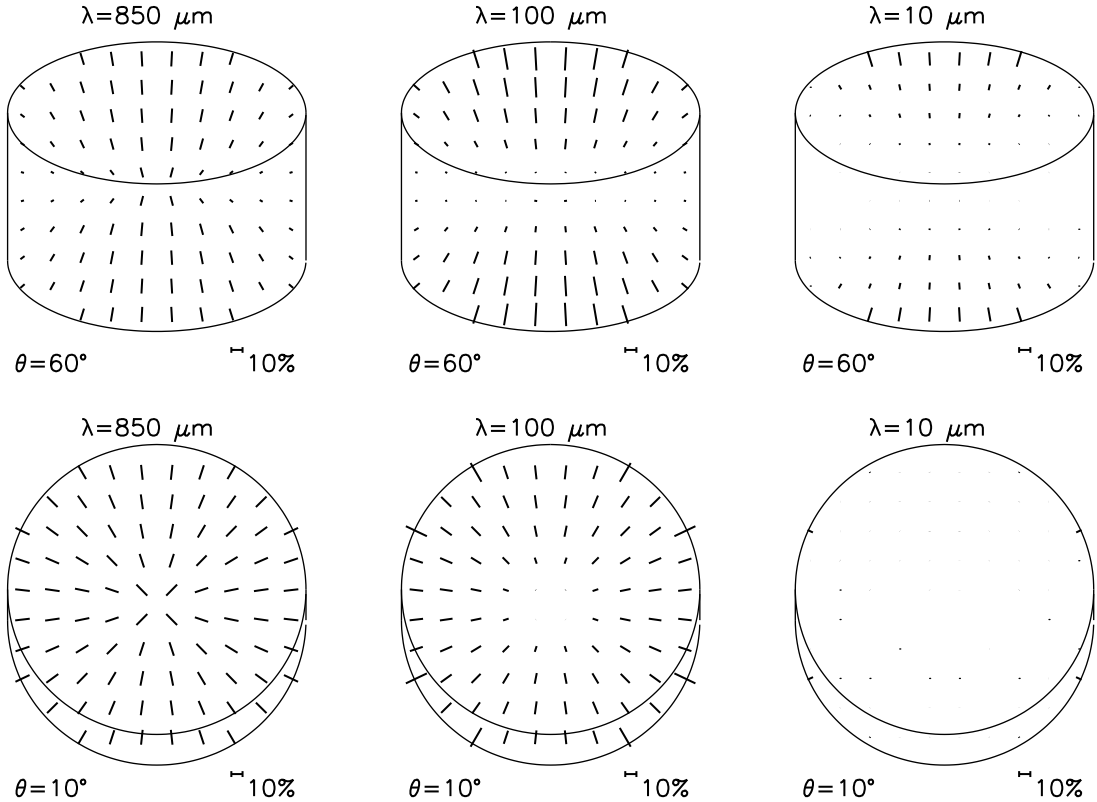}
\caption{
  Simulated observations.
Degree of polarization is calculated for
the radiation from the disk interior only.
\label{fig:visual_int}
}
\end{figure*}

Figure \ref{fig:visual}  
shows the effects of the disk inclination.
We calculate the polarized emission from the disk interior.
The viewing angle $\theta$ (=the angle of disk inclination) is the angle between the disk symmetry axis
and the line of sight.
We plot the direction of polarization for 3 different wavelengths and
2 different viewing angles.
The lines represent the direction of polarization.
Since we assume that magnetic field is azimuthal, the
direction of polarization is predominantly radial (see lower panels).
In Figure \ref{fig:visual_int} we show similar plots for
radiation from the disk interior only.
For $\lambda > 100 \mu m$, the polarization patterns in 
 Figure \ref{fig:visual_int} is very similar to those in
 Figure \ref{fig:visual}. But near the disk edges, Figure \ref{fig:visual}
 shows larger degree of polarization than Figure \ref{fig:visual_int}.
 This is because the emission from the disk interior is very weak there
 compared with that from the disk surface layer.
 For $\lambda < 100 \mu m$, the polarization patterns in 
 Figure \ref{fig:visual_int} are very different from
 those in Figure \ref{fig:visual}, because polarized emission from 
 the disk surface layer dominates that from the disk interior.
 Note that, 
since the degree of polarization of emission from the disk surface layer is
very sensitive to the maximum grain size in the surface layer,
the results for $\lambda < 100 \mu m$ should be very sensitive to
the maximum grain size in the surface layer.

\begin{figure*}[h!t]
\includegraphics[width=.88\textwidth]{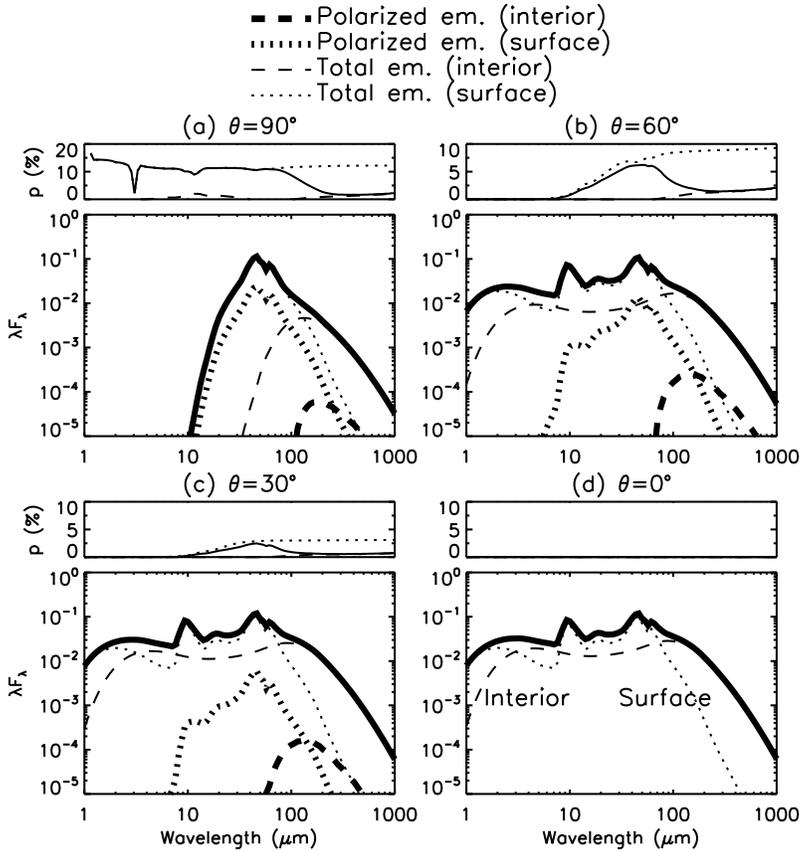}
\caption{
  Spectral energy distribution for
  four different viewing angles.
  When $\theta=90$ (i.e. for edge-on disk), 
 inner part of the disk (i.e. region
close to the star) is invisible because it is occulted by the 
outer part of the disk.
 Therefore the spectral energy distribution truncates
 for $\lambda < 10 \mu m$.
 When $\theta=0$ (i.e. for face-on disk),
 the polarized emission is zero as expected. 
 This is because the assumed magnetic field configuration is
 perfectly azimuthal. 
\label{fig:sedang}
}
\end{figure*}

While the polarimetry of the spatially resolved accretion disks is promising
with new generation of instruments (see \S 6.2), at present one can 
study disk magnetic fields with unresolved accretion disks. Below we provide
predictions for this case.
Figure \ref{fig:sedang} shows spectral energy
   distribution for such a disk for
  four different viewing angles.
  When $\theta=90$ (i.e. for edge-on disk), 
 inner part of the disk (i.e. region
close to the star) is invisible due to high opacity.
 Therefore the spectral energy distribution truncates
 for $\lambda < 10 \mu m$.
 When $\theta=0$ (i.e. for face-on disk),
 the polarized emission is zero as expected\footnote{
That is, we do not see thick dotted or thick dashed lines in 
Figure \ref{fig:sedang}(d).}.

\begin{figure*}[h!t]
\includegraphics[width=.48\textwidth]{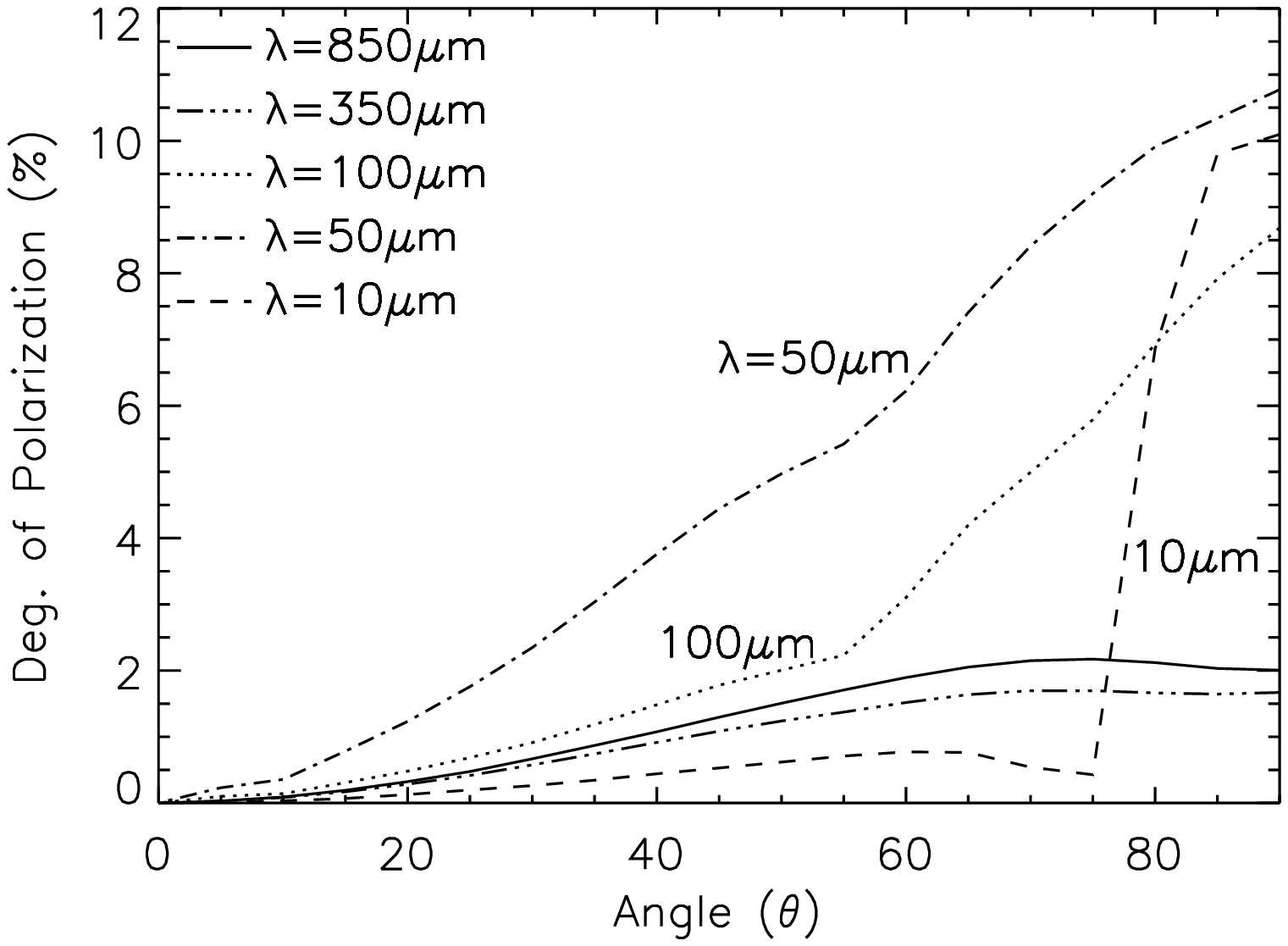}
\includegraphics[width=.48\textwidth]{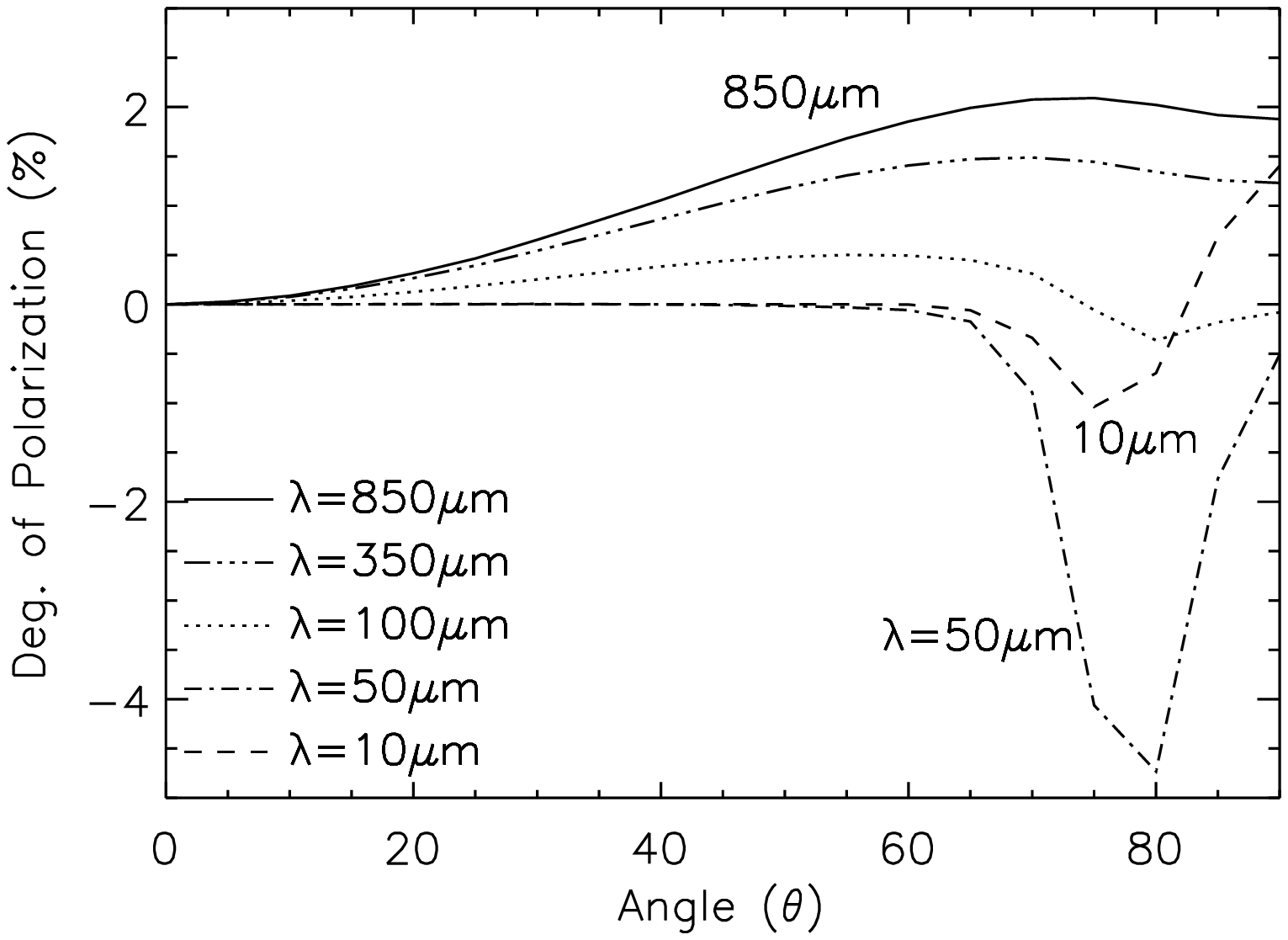}
\caption{
  Degree of polarization vs. viewing angle.
Left panel shows the degree of polarization for
total (i.e. interior + surface) emission, while the right panel shows that for
radiation from disk interior only.
\label{fig:tilt}
}
\end{figure*}

Finally, Figure \ref{fig:tilt} shows the change of 
the degree of polarization for selected wavelengths.
Left panel shows the degree of polarization for
total emission, while the right panel shows that for
radiation from the interior only.
The degree of polarization is large when the angle $\theta$ is small
(see left panel).
The sudden drop for $\lambda=10 \mu m$ is due to the following reason.
As the viewing angle drops, the
inner part of the disk suddenly becomes visible, which causes
a sudden increase of the
the total intensity (or flux).
But, the polarized intensity (or flux) does not change much
because the inner part of disk does not emit
polarized emission. Note that grains are not aligned
in the inner part of the disk.
For $\lambda=50 \mu m$, polarization is dominated by the disk surface
layers.

\section{Discussion}

\subsection{Other Alignment Processes}

In the paper above we concentrated on only one alignment process, namely,
on the RT mechanism. This mechanism has became so promising in explaining
of interstellar polarization partially because the main competitor, namely,
paramagnetic alignment mechanism (Davis-Greenstein 1951) and its later
modifications
were shown to have problems with aligning interstellar grains. In fact,
the fast spin-up mostly due to H$_2$ formation on the catalytic
sites over grain surface as suggested in Purcell (1979) is
a textbook process that is invoked to explain the efficient paramagnetic alignment.
Indeed, fast rotating grains should be immune to randomization by atomic 
bombardment and thus get aligned well.
However, the Purcell's spin-up was shown to 
be inefficient for most of grains in diffuse interstellar gas
due to thermal flipping of grains that was reported
in Lazarian \& Draine (1999ab). The thermal flips arise from the coupling
of vibrational and rotational degrees of freedom stemming from the
processes of internal relaxation within interstellar grains, in 
particular, Barnett (Purcell 1979) and nuclear relaxation (Lazarian \& Draine
1999b). As a result of flipping the direction of the Purcell's torques
acting on a grain alters and the grain gets ``thermally trapped''. It rotates
at a thermal velocity and therefore is subjected the randomization due to
random gaseous and ionic bombardment. 

The difference of the interstellar and disk grains is their size. The larger
grains in the disk are not thermally trapped. Therefore, potentially, the
processes of Purcell's spin-up are applicable. If temperatures of grains
and ambient gas are different, this may result in a spin-up that arises from
variations of the accommodation coefficient over grain surface (Purcell 1979),
provided that the temperatures of the gas and the grains in the disk differ.
The problem of such a scenario is that the paramagnetic alignment is slow, 
unless grains demonstrate enhanced magnetic susceptibility 
(e.g. are super-paramagnetic) (see Jones \& Spitzer 1967). Potentially, this
process that also aligns grains with long axes perpendicular
to magnetic field can enhance the alignment and therefore polarization.
However, we do not know about the abundance of the required grains within
the disks.  

For the largest grains, a particular mechanical alignment, which was
termed in Lazarian (1994) ``weathercock mechanism'' is applicable. In the
presence of gas-grain motions large irregular grains would tend to get aligned
with long dimension along the flow as their center of pressure and center of
mass do not coincide. However, the mechanism requires substantial relative
velocities of
gas and grain, which is not certain in the protostellar disks. Moreover,
we have showed above that very large grains do not produce polarized
radiation at the wavelength we deal with in this paper.

More promising may be the alignment of helical grains first mentioned
in Lazarian (1995) and discussed at some depth in Lazarian \& Hoang (2007). The mechanism is based on the interaction of an irregular grain with a flow of atoms. If some fraction of colliding atoms is not being absorbed by the grain surface due to grain accommodation coefficient not being equal to unity, the collisions with a flow of gaseous atoms should  cause 
the
alignment similar to that by anisotropic radiative torques. However,
as it is discussed in Lazarian (2007) we do not any compelling evidence 
for the operation of the mechanism in the studied astrophysical
environments. On the contrary, the alignment by radiative has been proven
to provide the observed alignment in a number of circumstances (see
Lazarian 2007). Thus we defer a quantitative discussion of the mechanical
alignment of helical grains. If the mechanism operates efficiently, 
it can only
increase the degree of alignment at the parts of the disk where the radiative torques start to fail, making polarization from aligned grains only
more important.      

All in all, while further studies of alternative alignment mechanisms seems
necessary, at present the discussed RT mechanism provides the safest bet.

\subsection{Observational Prospects}

Multifrequency observations of protostellar disks have become a booming field recently.
They have advanced substantially our knowledge of the disks and allowed theoretical
expectations to be tested. 

Magnetic fields are an essential components of the protostellar disks. They are likely
to be responsible to accretion (see Nomura 2002). Therefore observational studies of them are essential. In this respect our paper is the first, as far as we know, attempt to provide
the expectations of the polarization arising from accretion disks that is based on
the predictions of the grain alignment theory.   

Our study reveals that multifrequency polarimetry is very important for the protostellar
disks. The synthetic observations that we provide explicitly show that observations
at wavelength less than 100 $\mu$m mostly test magnetic fields of the skin layers, while
at longer wavelengths test magnetic fields of the bulk of the disk. Therefore polarimetry
can, for instance, test theories of accretion, e.g.
 layered accretion (Gammie 1996).

Combining the far-infrared polarimetry with polarimetric measurements at
different frequencies may provide additional insight into the magnetic
properties of protostellar accretion disks. For instance, circular polarization
arising from scattering of starlight from aligned grains (see Lazarian 2003) and
polarization in emission lines arising from the aligned atoms (see Yan \& Lazarian 2006, 2007)
can provide additional information about the magnetic in the outer parts and above the
accretions disks. 

Most of the present day polarimetry will be done for
not resolved 
protostellar disks.
The size of the 
T Tauri disks is usually less than
$\sim 300$ AU (see, for example, C01). 
If we take the distance to proto-stars to be around $\geq 100 pc$,
then the angular sizes of the disks are usually smaller than
$6\arcsec$. 
The angular resolution of SCUBA polarimeter (SCUPOL) is around
$14\arcsec$ (Greaves et al. 2000) and that of SHARC II polarimeter
(SHARP; Novak et al. 2004) at $350 \mu m$ is around $9\arcsec$.
Therefore it is not easy to obtain plots like Figures \ref{fig:visual}.
The angular resolution of the intended SOFIA polarimeter is 
around $5\arcsec$ at $53 \mu m$, 
$9\arcsec$ at $88 \mu m$, and 
$22\arcsec$ at $215 \mu m$.
We see that the intended SOFIA polarimeter will be
 at the edge of resolving structure of close-by disks, while other instruments
will not resolve typical T Tauri disk. Therefore for most
of the near future observations
our predictions in Fig.~13 and 14 are most relevant.
  
 Higher  resolution
polarimetry is expected in future, however. This will make our predictions
of polarization the resolved accretion disks in Fig.~11 and 12 testable.
Note, that the actual structure of magnetic fields may be much more complex
than the one in our simple model. 

While in the paper we dealt with the protostellar accretion disks, our results
are suggestive of importance of polarimetric
studies of magnetic fields in the disks of evolved stars. In a broader context, the
present paper is one of the first 
studies that make use of the advances in grain alignment theory
to extend the utility of polarimetric studies of magnetic field beyond its traditional
interstellar domain.

\subsection{Effects of scattering}
In this paper, we do not consider the effects of scattering.
Below we show that effects of scattering is indeed
less important
for emissions from disk interior in the FIR regime.
We will provide more detailed calculations on the effects of
scattering for shorter wavelengths elsewhere.

\begin{figure*}[h!t]
\includegraphics[width=.48\textwidth]{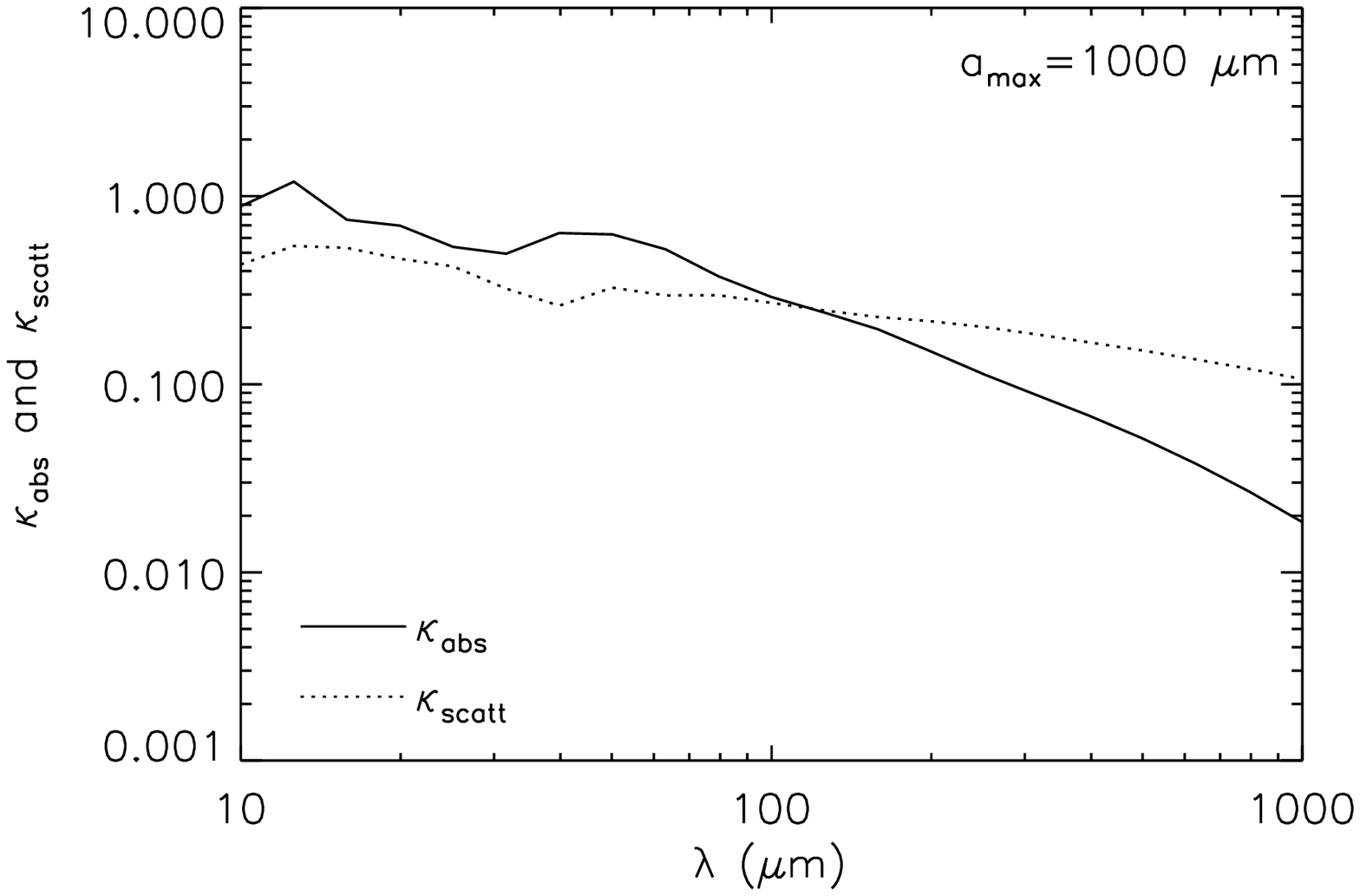}
\includegraphics[width=.48\textwidth]{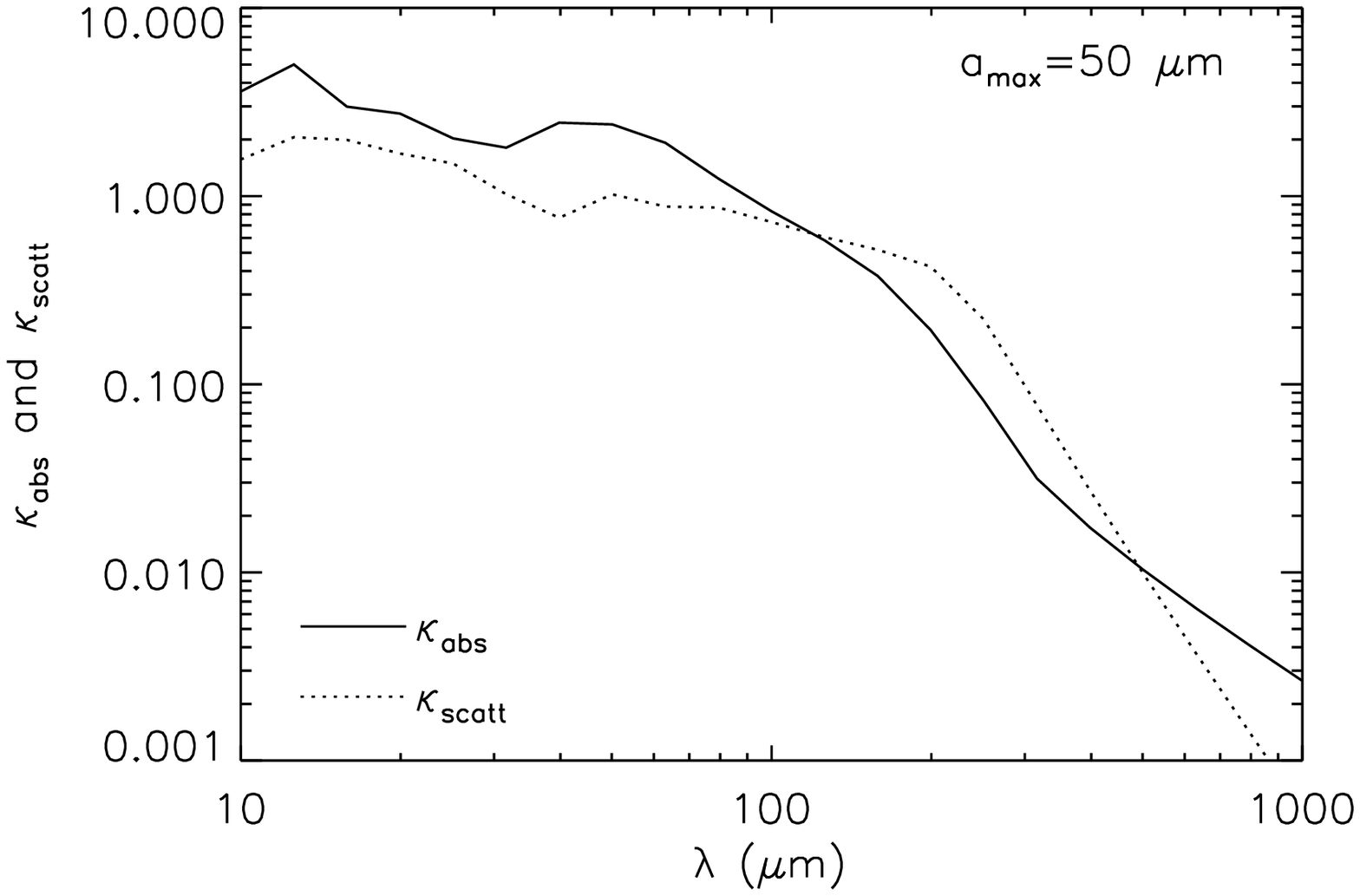}
\caption{
Mass absorption and scattering coefficients.
Y-axes are in
arbitrary units.
{\it Left panel:} The maximum grain size, $a_{max}$, is 1000$\mu$m.
             The scattering coefficient is larger when 
             $\lambda > 100 \mu$m.
{\it Right panel:} The maximum grain size, $a_{max}$, is 50$\mu$m.
             The scattering coefficient is similar to or less than
             the absorption coefficient.
In both cases, the minimum grain size is 0.01$\mu$m.
\label{fig:sca-abs}
}
\end{figure*}

\begin{figure*}[h!t]
\includegraphics[width=.48\textwidth]{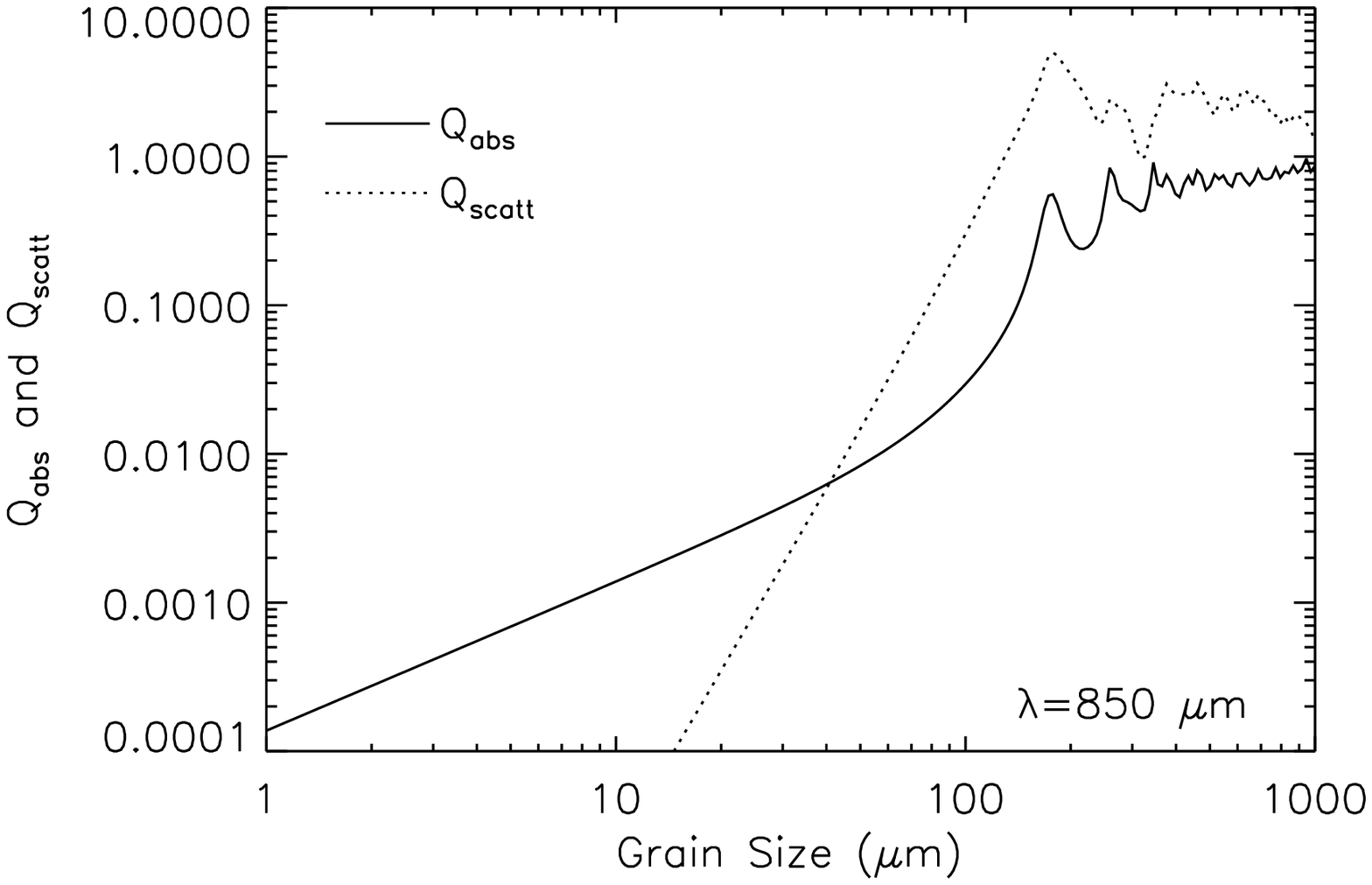}
\includegraphics[width=.48\textwidth]{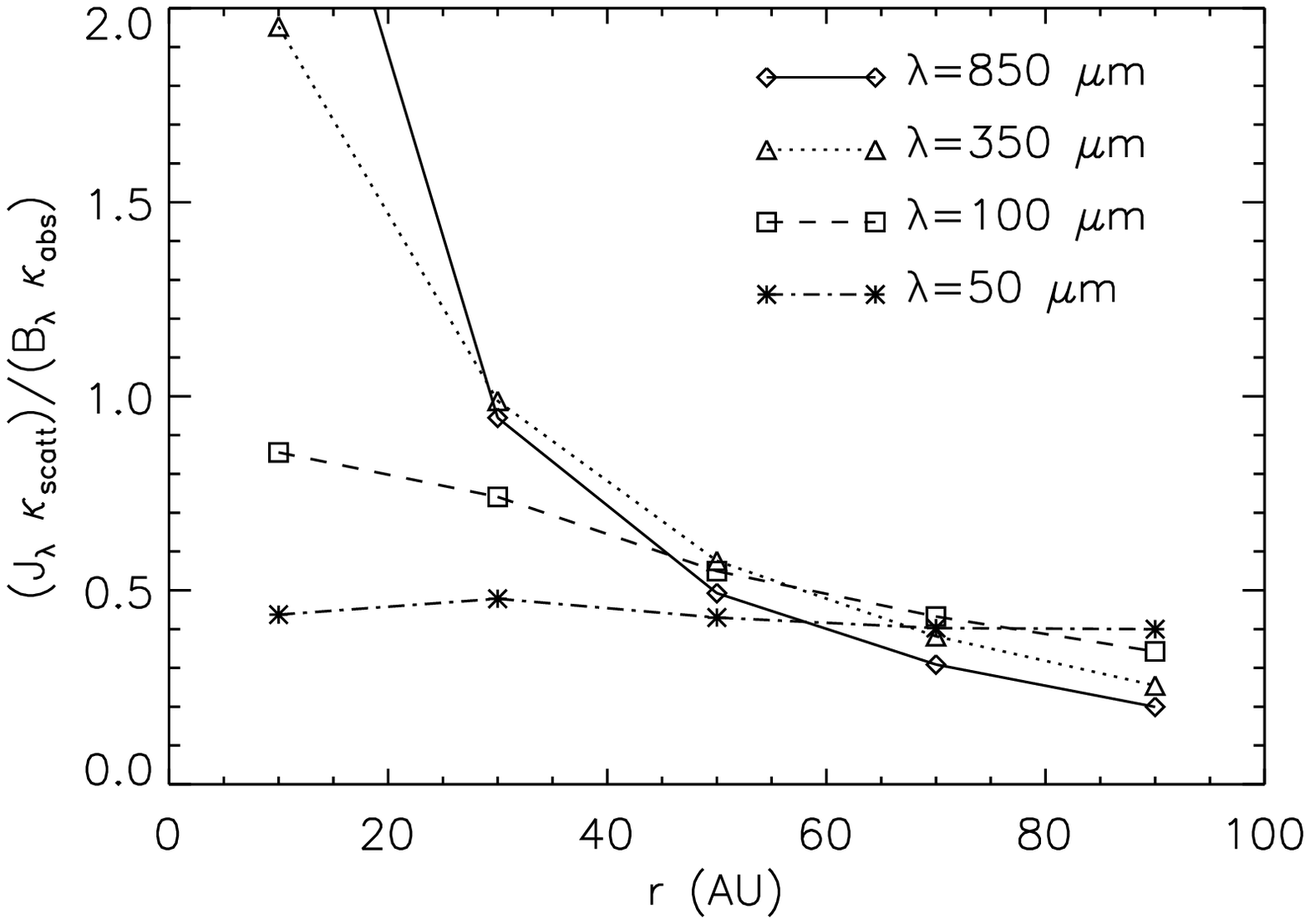}
\caption{Scattering and absorption 
         coefficients.
         The wavelength is fixed for $\lambda=850\mu$m.
         The grains are spherical silicate coated with water ice.
         The scattering coefficient is larger than the absorption
         one for large grains.
\label{fig:fixedl}
}         
\caption{Comparison between polarization by scattering and emission.
   When the value in y-axis is less than 1, we expect that    
   absorption is more important for polarization.
   The values are calculated in the mid-plane of the disk.
   When we calculate the values at locations above the mid-plane
   the values may be smaller.
\label{fig:qxj}
}
\end{figure*}

In the case $2\pi a/\lambda \ll 1$ we can assume that $Q_{scatt}$ is
negligibly smaller than $Q_{abs}$. However, since we are dealing with large grains here, we cannot simply assume the inequality. 
We first compare the relative magnitudes of mass scattering
and absorption coefficients, $\kappa_{scatt}$ and $\kappa_{abs}$,
when grain size distribution follows an MRN-type power law,
$dN \propto a^{-3.5}  da$,
between $a_{min}=0.01 \mu$m and $a_{max}$ (see also section
2.1). Figure~\ref{fig:sca-abs} shows
$\kappa_{scatt}$ is larger than $\kappa_{abs}$ for
$\lambda \geq 100\mu$m in case $a_{max} = 1000\mu$m. 
This is because the scattering 
efficiency is larger than the absorption one when
$\lambda \sim 2\pi a$ 
(Fig.~\ref{fig:fixedl}).
This trend is generally observed when $\lambda \geq 100\mu$m. 
When the cutoff size, $a_{max}$, gets smaller, $\kappa_{scatt}$ becomes
subdominant (see the right panel of Fig.~\ref{fig:sca-abs}). 

Polarization by scattering is proportional to 
$\sim J_{\lambda} \kappa_{scatt}$, where $J_{\lambda}$ is the mean radiation field, while that by emission is proportional
$\sim B_{\lambda} \kappa_{abs}$, where $B_{\lambda}$ is the intensity
of the blackbody radiation at the point of interest.
Figure \ref{fig:qxj} shows this $J\kappa_{scatt}$ to $B\kappa_{abs}$ ratio at selected points on the
disk mid-plane. We use the disk model in C01 to
calculate $J_{\lambda}$ and $B_{\lambda}$.
We only include emission from disk interior.
Recall that most FIR emission from disk interior are from 
$r > 10$AU (see Fig.~\ref{fig:red}). 
Therefore, the $J\kappa_{scatt}$ to $B\kappa_{abs}$ ratio for $r > 10$AU 
concerns us most.
This crude estimate tells us that polarization by scattering is less important than that by
emission in the FIR regime. 
Nevertheless, readers should keep in mind that our polarization 
calculations in this paper reflect only the emitted component
and, therefore, the radiative transfer adopted is approximate.

\begin{figure*}[h!t]
\includegraphics[width=.48\textwidth]{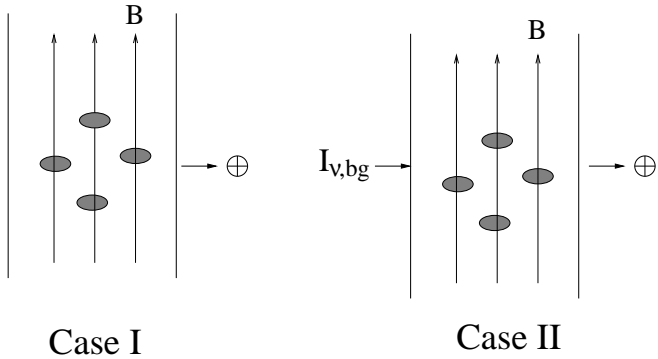}
\caption{In Case I, all parts of the disk along the line of sight
         have aligned grains. In this case,
         the direction of polarization is perpendicular 
         to the magnetic field.
         In Case II, the aligned part lies in front of
         unaligned part. In this case,
         the direction of polarization depends on the
         relative strength of the local source function and 
         the background radiation.
         Note that $I_{\nu,bg}$ is unpolarized radiation from background
         regions of the disk.
\label{fig:f18}
}         
\end{figure*}

How important is the self scattering effect?
Let us first consider the case that the entire disk volume has aligned grains (Case I
in Fig. \ref{fig:f18}). 
As we will see right below, self scattering/absorption does not change 
the direction of polarization in Case I in Fig. \ref{fig:f18}.
The observed radiation parallel to
the magnetic field becomes
\begin{eqnarray}
I_{\lambda,\|} = \int B_{\lambda} 
                          \exp(-\tau_{ext,\|}) d\tau_{abs,\|} 
    \nonumber \\
   = \frac{1}{1+\alpha_{\|}}\int B_{\lambda} \exp(-\tau_{ext,\|}) 
  d(\tau_{ext,\|})=\frac{B_{\lambda}}{1+\alpha_{\|}}[1-\exp(-\tau_{ext,\|})],
\end{eqnarray}  
where $\tau_{ext,\|}\equiv\tau_{abs,\|}+\tau_{scatt,\|}$ 
and we assume that the gas is uniform 
and that 
$\alpha_{\|} \equiv  \tau_{scatt,\|}/\tau_{abs,\|} $
does not change along the line of sight. 
For simplicity, we ignore the radiation that scatters into the line of sight.
We will have a similar expression for the perpendicular direction.
Therefore, 
     we have
\begin{equation}
 I_{\perp}-I_{\|} \propto
 \frac{ 1-\exp(-\tau_{ext,\perp}) }{ 1+\alpha_{\perp} }- 
 \frac{ 1-\exp(-\tau_{ext,\|}) }{ 1+\alpha_{\|} },
\end{equation}
which is positive when $\tau_{ext,\perp} > \tau_{ext,\|}$
and $\alpha_{\perp} \approx \alpha_{\|}$.
This means that the direction of polarization is perpendicular to
the magnetic field.
Recall that most of the outer part of disk (i.e. $r>10$AU) has aligned grains 
(see Figure \ref{fig:align_int}).
Therefore, the situation is similar to Case I and 
we expect that including self scattering does not change our qualitative
results in earlier sections.

In Case II, however, self scattering can be potentially important.
Case II will happen when we have a slab of aligned grains in front of
unaligned grains (see Fig. \ref{fig:f18}). 
The observed intensity becomes
\begin{equation}
  I_{\lambda,\|} = I_{\lambda,bg} \exp(-\tau_{ext,\|})
           +\frac{B_{\lambda}}{1+\alpha_{\|}}
            [1-\exp(-\tau_{ext,\|})]
\end{equation}
for the direction parallel to the magnetic field, and
\begin{equation}
  I_{\lambda,\perp} = I_{\lambda,bg} \exp(-\tau_{ext,\perp})
           +\frac{B_{\lambda}}{1+\alpha_{\perp}}
            [1-\exp(-\tau_{ext,\perp})]
\end{equation}
for the direction perpendicular to the magnetic field.
Here, for simplicity, we assume the medium is uniform.
In optically thin case, we have
\begin{equation}
  I_{\lambda,\|} = I_{\lambda,bg} - I_{\lambda,bg} \tau_{ext,\|} 
           +\frac{B_{\lambda}}{1+\alpha_{\|}}
            \tau_{ext,\|}
          =  I_{\lambda,bg} - I_{\lambda,bg} \tau_{ext,\|}
           +B_{\lambda}
            \tau_{abs,\|}
\end{equation}
for the direction parallel to the magnetic field, and
\begin{equation}
  I_{\lambda,\perp} = I_{\lambda,bg} - I_{\lambda,bg} \tau_{ext,\perp}
           +\frac{B_{\lambda}}{1+\alpha_{\perp}}
            \tau_{ext,\perp}
         = I_{\lambda,bg} - I_{\lambda,bg} \tau_{ext,\perp}
           +B_{\lambda}
            \tau_{abs,\perp}
\end{equation}
for the direction perpendicular to the magnetic field.
Then we have
\begin{eqnarray}
 I_{\perp}-I_{\|} \propto 
  B_{\lambda} (\tau_{abs,\perp}-\tau_{abs,\|} )
 - I_{\lambda,bg} ( \tau_{ext,\perp}-\tau_{ext,\|} ) \nonumber \\ 
  =\left[ B_{\lambda} 
 - I_{\lambda,bg} 
   \left( 1+ \frac{ \tau_{scatt,\perp}-\tau_{scatt,\|}  }
             { \tau_{abs,\perp}-\tau_{abs,\|} } \right)
  \right]
   (\tau_{abs,\perp}-\tau_{abs,\|} ).
\end{eqnarray}
Therefore, the direction of polarization can change as a result of
scattering.

\begin{figure*}[h!t]
\includegraphics[width=.48\textwidth]{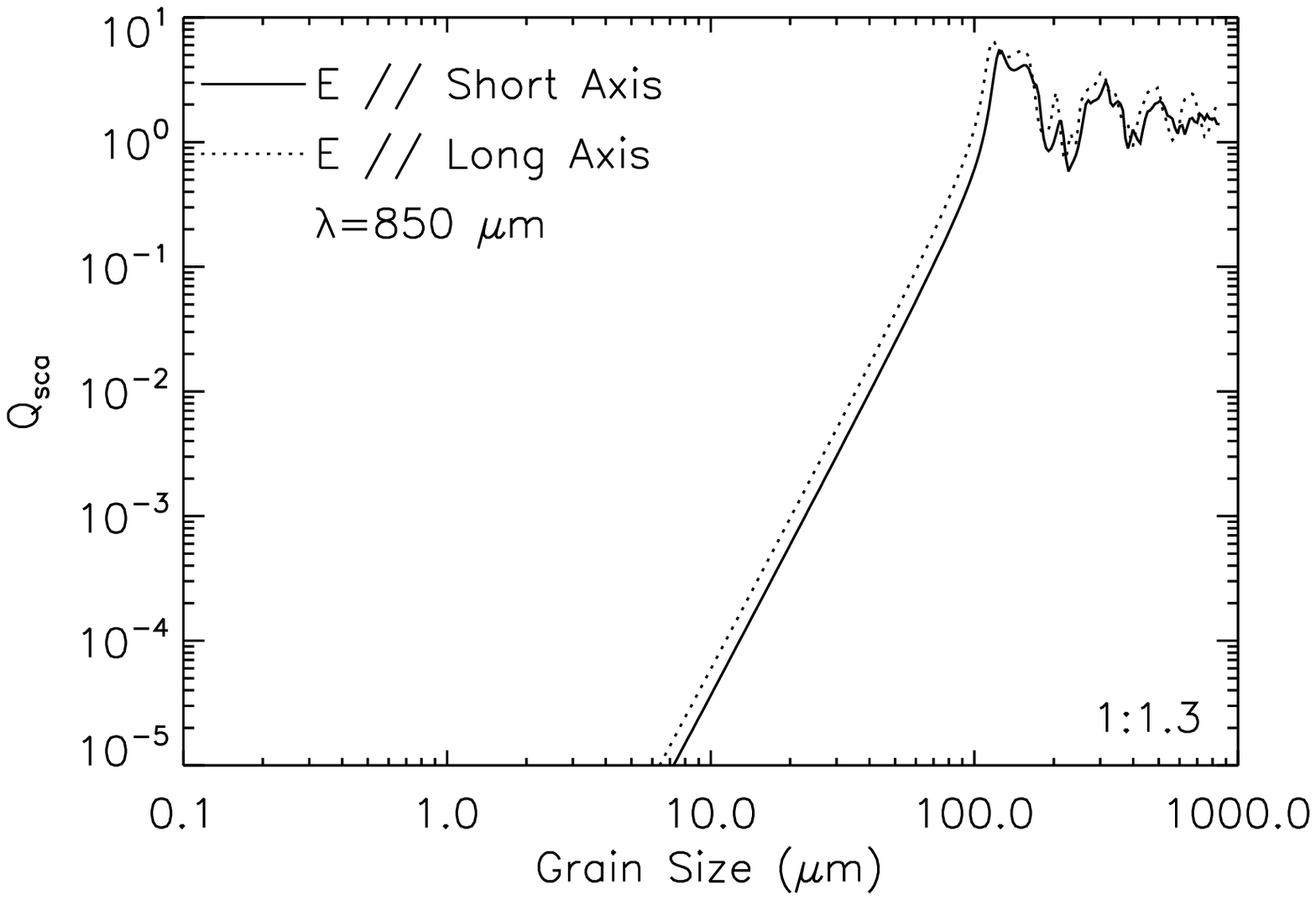}
\includegraphics[width=.48\textwidth]{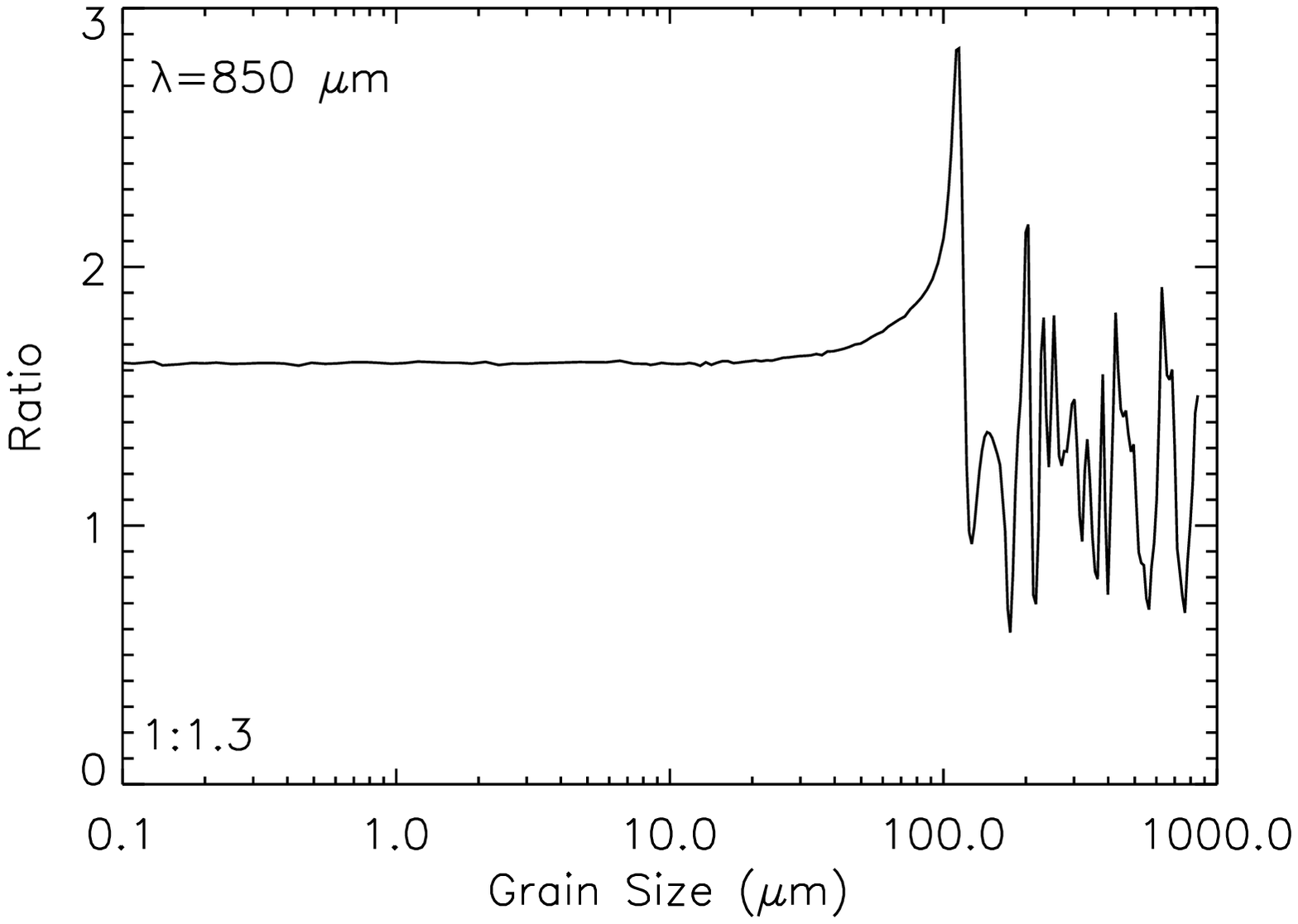}
\caption{Scattering efficiency vs. grain size. 
  When grains are oblate spheroid,
  the scattering cross-section depends on the direction of polarization. 
  similar to that of absorption (see Fig. 5). 
  See the caption of Figure 5 for details.
\label{fig:f17}
}         
\end{figure*}

The ratio $( \tau_{scatt,\perp}-\tau_{scatt,\|}  )/
             ( \tau_{abs,\perp}-\tau_{abs,\|} )$
cannot be very large.
Fig. \ref{fig:f17} implies that scattering does not cause
polarization in the geometrical optics regime (i.e. $\lambda/2 \pi <  a$).
On the other hand, Fig. \ref{fig:fixedl} shows that
scattering is negligible when $a$ is a few time smaller than $(\lambda/2\pi)$.
This means that scattering dominates polarization only for a limited range: $\beta (\lambda/2\pi) < a < (\lambda/2\pi)$, 
where $\beta \approx 0.2 - 0.3$.
However, scattering dominates absorption for  
$\beta (\lambda/2\pi) < a$ (see Fig. \ref{fig:fixedl}).
Therefore, we expect that
\begin{equation}
   \frac{ \tau_{scatt,\perp}-\tau_{scatt,\|}  }        
         { \tau_{abs,\perp}-\tau_{abs,\|} }  
 <   \frac{ \tau_{scatt} } 
     { \tau_{abs}  }
\end{equation} 
(see Fig. \ref{fig:sca-abs} for the ratio $\kappa_{scatt} /
      \kappa_{abs}$).


\section{Summary}

Making use of the recent advances in grain alignment theory
we calculated grain alignment by RTs in a magnetized T Tauri disk.
Based on this, we calculated polarized emission from the disk.
Our results show that
\begin{itemize}

\item Polarization arising from aligned grains reveals magnetic fields
of the T Tauri disk. 

\item Grain size distribution is the most important factor
that determine the degree of polarization. 

\item Disk interior dominates polarized emission in FIR/sub-millimeter
wavelengths. When there are many grains with maximum grain size of
 $\sim 1000 \mu m$,
the degree of polarization is around or less than
$\sim 2$\% in these wavelengths.
However, when the maximum grains are smaller, we expect higher
degree of polarization.

\item Disk surface layer dominates polarized emission in mid-IR wavelengths.
The degree of polarization is very sensitive to the
maximum size of grain in the disk surface layer.
When the maximum grain size is as large as $\sim 1 \mu m$, we expect
$\sim 10$\% of polarization at $\lambda \sim 50 \mu m$.
However, when the maximum size is smaller the the value will drop.

\item Our study of the effect of the disk inclination predicts substantial
changes of the degree of polarization with the viewing angle.
The coming mid-IR/FIR polarimeters are very promising for studies of magnetic
fields in protostellar disks.
 
\end{itemize}

\acknowledgements
Jungyeon Cho's work was supported by the Korea Research Foundation grant
funded by the Korean Government (KRF-2006-331-C00136). A. Lazarian acknowledges the support by the NSF grants AST 02 43156 and AST 0507164, as well as by the
NSF Center for Magnetic Self-Organization in Laboratory and Astrophysical 
Plasmas. 
The work of Jungyeon Cho was also supported in part by
Korea Foundation for International Cooperation of Science \&
Technology (KICOS) through the Cavendish-KAIST Research Cooperation 
Center.


\begin{thebibliography}{8.}
\addcontentsline{toc}{section}{References}
\bibitem[]{1288} Aitken, K., Efstathiou, A., McCall, A., \&
          Hough, J. 2002, MNRAS, 329, 647

\bibitem[]{1291} Balbus, S. A., \& Hawley, J. F. 1991, ApJ, 376, 214
\bibitem[]{1292} Bethell, T., Chepurnov, A., Lazarian, A. \& Kim, J. 2007, ApJ, in press

\bibitem[]{1293} Bohren, C.F. \& Huffman, D.R. 1983, {\it 
     Absorption and Scattering of Light by Small Particles}
     (New York: Wiley)

\bibitem[]{1297} Chandrasekhar, C. 1961, in {\it Hydrodynamic and 
     Hydromagnetic Stability} (Oxford University Press, Oxford)

\bibitem[]{1300} Chiang, E. \& Goldreich, P. 1997, ApJ, 490, 368 (CG97)
\bibitem[]{1301} Chiang, E. \& Goldreich, P. 1999, ApJ, 519, 279
\bibitem[]{1302} Chiang, E.,  2001, ApJ, 547, 1077 (C01)
\bibitem[]{1303} Cho, J. \& Lazarian, A. 2005, ApJ, 631, 361
\bibitem[]{1304} Dolginov A.Z. 1972, Ap\&SS, 16, 337
\bibitem[]{1305} Dolginov A.Z. \& Mytrophanov, I.G. 1976, Ap\&SS, 43, 291 

\bibitem[]{1307} Draine, B. 1985, ApJS, 57, 587
\bibitem[]{} Draine, B. \& Flatau, P. 1994, J. Opt. Soc. Am. A, 11, 1491
\bibitem[]{1308} Draine, B. \& Lee, H. 1984, ApJ, 285, 89
\bibitem[]{1309} Draine, B. \& Lazarian, A. 1998, ApJ, 508, 157
\bibitem[]{1309} Draine, B. \& Weingartner, J. 1996, ApJ, 470, 551 (DW96)
\bibitem[]{1310} Draine, B., \& Weingartner, J. 1997, ApJ, 480, 633 (DW97)
\bibitem[]{1311} Gammie, C. 1996, ApJ, 462, 725
\bibitem[]{1312} Greaves, J., Holland, W., Jenness, T., \& Hawarden, T.
           2000, Nature, 404, 732
\bibitem[]{1314} Greenberg, M. 1968, in 
            {\it Stars and Stellar Systems}, Vol. 7, 
            {\it Nebulae and Interstellar Matter}, eds. by
            B. M. Middlehurst \& L. H. Aller, 
            (Univ. of Chicago Press, Chicago), p. 221
\bibitem[]{1320} Hall, J. 1949, Science, 109, 166
\bibitem[]{1321} Hildebrand, R., \& Dragovan, M. 1995, ApJ, 450, 663
\bibitem[]{1322} Hiltner, W. 1949, Science, 109, 165
\bibitem[]{1323} Hoang, T. \& Lazarian, A. 2007, ApJ, submitted

\bibitem[]{1325} Laor, A. \& Draine, B. 1993, ApJ, 402, 441
\bibitem[]{1326} Lazarian, A. 1994, Ap\& SS, 216, 235
\bibitem[]{13261} Lazarian, A. 1995, MNRAS, 216, 235
\bibitem[]{1327} Lazarian, A. 2003, Journal of Quantitative Spectroscopy 
            and Radiative Transfer, 79, 881
\bibitem[]{13271} Lazarian, A. 2007, Journal of Quantitative Spectroscopy 
            and Radiative Transfer, 106, 225
\bibitem[]{1329} Lazarian, A., \& Draine, B. 1999a, ApJ, 516, L37
\bibitem[]{1330} Lazarian, A., \& Draine, B. 1999b, ApJ, 520, L67
\bibitem[]{1331} Lazarian, A. \& Hoang, T. 2007, MNRAS, in press

\bibitem[]{1333} Lee, H. \& Draine, B. 1985, ApJ, 290, 21
\bibitem[]{1334} Mathis, J., Rumpl, W., \& Nordsieck, K. 
           1977, ApJ, 217, 425 (MRN)
\bibitem[]{1336} Nomura, H. 2002, ApJ, 567, 587
\bibitem[]{1337} Novak, G. et al. 2004, in {\it Millimeter and
           Submillimeter Detectors for Astronomy II}, eds.
           J. Antebi \& D. Lemke, Proceedings of the SPIE, Vol. 5498, p. 278
\bibitem[]{1340} Purcell, E. 1979, ApJ, 231, 404
\bibitem[]{} Roberge, W. \& Lazarian, A. 1999, MNRAS, 305, 615
\bibitem[]{1341} Tamura, M., Hough, J., Greaves, J., Morino, J.-I.,
           Chrysostomou, A., Holland, W., \& Momose, M. 1999,
           ApJ, 525, 832

\bibitem[]{1345} Spitzer, L., \& McGlynn, T. 1979, ApJ, 231, 417
\bibitem[]{1346} Velikov, S. 1959, J. Expl. Theoret. Phys. (USSR), 36, 1398
\bibitem[]{1347} Weingartner, J. \& Draine, B. 2001, ApJ, 548, 296
\bibitem[]{1348} Weingartner, J., \& Draine, B. 2003, ApJ, 589, 289
\bibitem[]{1349} Yan, H. \& Lazarian, A. 2006, ApJ, 653, 1292
\bibitem[]{1350} Yan, H. \& Lazarian, A. 2007, ApJ, 657, 618

\end{thebibliography}
\end{document}